\documentclass[epj]{svjour}

\usepackage{microtype}
\usepackage[pdfborder={0 0 0}]{hyperref} 
\usepackage{mathtools}
\usepackage{amstext}
\usepackage{amsmath}
\usepackage{amssymb}
\usepackage{amsfonts}
\usepackage{graphicx}
\usepackage{color}
\usepackage{cite}
\usepackage[normalem]{ulem} 

\usepackage[version=4]{mhchem}

\renewcommand{\vec}[1]{\boldsymbol{#1}}

\definecolor{Red}{rgb}{0.9,0.0,0.1}
\definecolor{Lila}{rgb}{0.7,0.0,0.9}
\definecolor{Darkblue}{rgb}{0.22,0.33,0.64}
\definecolor{Darkgray}{rgb}{0.4,0.4,0.4}
\definecolor{Blue}{rgb}{0.1,0.0,0.9}

\begin{document}

\title{Surfactant-loaded capsules as Marangoni microswimmers
   at the air-water interface:
  Symmetry breaking and spontaneous propulsion  by surfactant
    diffusion and advection}
  \author{Hendrik Ender\inst{1}, Ann-Kathrin Froin\inst{2}, Heinz Rehage\inst{2}
    \and Jan Kierfeld\inst{1}}
\institute{                    
   Department of Physics, Technische Universit\"{a}t Dortmund, 
  44221 Dortmund, Germany \and
   Department of Chemistry and Chemical Biology,
  Technische Universit\"{a}t Dortmund, 
  44221 Dortmund, Germany
}
\date{Received: date / Revised version: date}

\authorrunning{H. Ender, A.-K. Froin, H. Rehage, and  J. Kierfeld}
\titlerunning{Surfactant-loaded capsules as Marangoni microswimmers}

\abstract{
We present a realization of a fast  interfacial Marangoni microswimmer
by a   half-spherical  alginate capsule at the air-water
interface, which diffusively releases water-soluble spreading molecules
(weak surfactants such as  polyethylene glycol (PEG)),
which act as ``fuel'' by   modulating the
    air-water interfacial tension.
    For a number of different fuels, 
    we can observe symmetry breaking and spontaneous propulsion
    although the alginate particle and emission are isotropic.
    The  propulsion mechanism is similar to soap or camphor boats,
    which  are, however, typically  asymmetric in shape or emission
    to select a swimming direction. 
    We develop a  theory of Marangoni boat propulsion
     starting from  low Reynolds numbers by analyzing
  the coupled problems of surfactant
  diffusion and advection and  fluid flow, which includes 
  surfactant-induced fluid Marangoni flow, and surfactant adsorption at the
  air-water interface;
  we also include a possible evaporation of surfactant.
  The swimming velocity is determined by the balance of drag and
  Marangoni forces. 
  We show  that
  spontaneous symmetry breaking resulting in propulsion
  is  possible above a critical dimensionless surfactant  emission rate 
   (Peclet number).
   We  derive the relation  between Peclet number and swimming speed
     and generalize to higher 
     Reynolds numbers utilizing  the concept of the Nusselt number.
     The theory  explains the observed swimming speeds for PEG-alginate
     capsules,  and we
     unravel the differences to other Marangoni boat systems
     based on camphor, which are mainly caused by  
   surfactant evaporation from  the liquid-air interface.
   The capsule Marangoni microswimmers also exhibit 
   surfactant-mediated repulsive interactions with walls, which can
   be qualitatively explained by surfactant accumulation at the wall.
%
%
 } 

\maketitle

\section{Introduction}

Designing and understanding 
self-propelling  biological or  artificial microswimmers
is the basis for the physics of active systems.
Swimming on the microscale is governed by low Reynolds numbers
and requires special propulsion
mechanisms which are effective in the presence of dominating viscous
forces.
The first propulsion principle that comes to mind is
shape-changing swimmers, which
deform their body in a cyclic way in order to propel.
At low Reynolds numbers,
the cyclic deformation pattern of a swimmer
must not be invariant under time-reversal
according to the scallop theorem \cite{Purcell1977}. 
In nature, many
different examples of deformation swimmers can be found such as  bacteria,
algae and spermatozoa \cite{Lauga2009}.
Realizing this concept in synthetic microswimmers is often difficult
as the scallop theorem requires control of
at least two  parameters.

Shape changing swimmers force the surrounding fluid via
no-slip boundary conditions on the surface of their moving parts. 
Another successful class of synthetic microswimmers
is phoretic swimmers, which actively create slip
velocities at their surface. 
Self-propelling
phoretic swimmers autonomously
create gradients in external fields such as
concentration of a ``fuel'' or temperature, which in turn give rise
to symmetry-breaking interfacial fluid flow in a thin interaction layer
\cite{Anderson1989}.
This fluid flow constitutes an effective slip velocity 
 leading to propulsion
\cite{ebbens2010,Illien2017}. Examples of such autophoretic swimmers
are thermophoretic or diffusiophoretic swimmers, which
generate gradients in temperature or concentration of interacting particles
along their body.
Self-diffusiophoretic swimmers generate
a non-vanishing interfacial slip velocity on the particle surface via
asymmetries in the  solute concentration field and a short-range interaction
between solute and swimmer \cite{Anderson1989}.
Diffusiophoretic
models typically neglect advection of the fuel concentration
\cite{Popescu2009,Popescu2011,Mozaffari2016}, but this has been included
in Refs.\ \cite{Michelin2014,Yariv2015}.
A lot of different aspects of swimmer behavior have been studied
for self-diffusiophoretic swimmers such as 
efficiency  \cite{Sabass2010},
confinement effects \cite{Popescu2009,Mozaffari2016} cargo transport
\cite{Popescu2011,Baraban2012} or
the rich behavior during collisions with walls
\cite{Uspal2015,Bayati2019}.

While diffusiophoresis creates concentration gradients within
the liquid surrounding the swimmer, 
concentration gradients or  surface active molecules (surfactants)
within the  interface of \emph{liquid} swimmers
can also generate symmetry-breaking interfacial
forces based on the Marangoni effect \cite{Scriven1960}.
These propulsion mechanism based on the Marangoni effects
are utilized in different  liquid Marangoni swimmers, such as 
active liquid droplets or active emulsions \cite{Herminghaus2014}.
Examples are pure water in an
 oil-surfactant medium (squalane and monoolein) \cite{Izri2014}
 or liquid crystal droplets in surfactant solutions
 \cite{Herminghaus2014} but many
 other systems can be generated making this a versatile route
 to microswimmer production. 
 This type of Marangoni swimmer is a liquid drop fully immersed in a
 liquid carrying surfactant,
 and propulsion is generated by the Marangoni effect
 along the liquid-liquid interface between swimmer and surrounding liquid,
 where a surfactant concentration gradient is maintained. 
 In Ref.\ \cite{Izri2014}, an auto-diffusiophoretic  mechanism
 \cite{michelin2013,Michelin2014} 
 has been proposed to maintain the surfactant concentration gradient.
 Another  mechanism that has been proposed
 is  increased adsorption of surfactant  at the
 front (in swimming direction)
 of the swimmer, which depresses the interfacial tension
 in the front \cite{Yoshinaga2012,Herminghaus2014,Schmitt2016}.
 This gives rise to a Marangoni stress 
 toward the  back (where the interfacial tension is higher).
 The Marangoni stress forces the surrounding fluid toward the back
 of the swimmer  resulting in a swimmer motion
 toward the front of the swimmer.
 For all proposed mechanisms,
 the liquid swimmer autonomously maintains an increased surfactant
 concentration in the front  of its interface with the surrounding liquid,
 and it propels in the direction of \emph{higher} surfactant
 concentration at its own interface.

 The self-phoretic and Marangoni
swimming mechanisms discussed so far do not generate net forces
on the swimmer
but non-vanishing slip velocities on the particle surface via
asymmetries in a temperature or solute concentration field.
There is another
class of self-propelling swimmers partly based on the Marangoni effect
and  with a long history \cite{Tomlinson1864}, 
which are so-called soap or camphor boats (or surfers), which we call
\emph{Marangoni boats} in the following.
Marangoni boats are moving at the liquid-air interface \cite{Nakata2015}; 
typically, they are \emph{solid} swimmers and
operate at the centimeter scale.
They are often used
as a popular demonstration experiment for the Marangoni effect
 \cite{Renney2013}.
 As ``fuel'' serve surface active molecules, which are deposited on
 the floating swimmer \cite{Renney2013}
 or  in which the swimmer is soaked
 \cite{Hayashima2001,Nagayama2004,
   Soh2008,Akella2018,Boniface2019,Sur2019}, or the swimmer body itself
 is made from dissolving surfactant \cite{Loffler2019}. 
 There are many examples based on DMF (dimethylformamide) \cite{Wang2016},
 alcohol \cite{Renney2013,Sur2019}, soap
 \cite{Sur2019}, camphor
 \cite{Hayashima2001,Nagayama2004,Soh2008,Suematsu2014,Akella2018,Boniface2019}
 or camphene \cite{Loffler2019}
that have also been characterized quantitatively. 
The surfactant molecules are emitted or dissolved from the swimmer
and a radial concentration gradient is established at the air-water interface
by diffusion, eventually aided by evaporation for volatile surfactants.
The radial concentration gradient creates (i) surface tension
gradients and (ii) Marangoni stresses on the fluid.
This leads, however, not necessarily to swimming as long as the 
surface tension  is symmetric and uniform around the swimmer.
The surface tension is pulling in normal direction on the
closed  air-water-swimmer contact line.
A  uniform surface tension cancels along any arbitrarily
shaped closed three-phase contact line, but a  gradient in
surfactant concentration along the contact line can generate
a net propulsion force.
We call this net force generated by surface tension gradients 
  \emph{direct Marangoni force} in the following.
Also symmetry-broken Marangoni flows  created by
the Marangoni effect  can contribute to (or impede)
the propulsion 
  via hydrodynamic drag onto the swimmer surface. We denote the resulting
  forces that Marangoni flows exert by \emph{Marangoni flow
    forces} in the following.
 The Marangoni boat mechanism is relying on
 both types of forces.
If surfactant 
emission is anisotropic the boat is, in general,
propelled into the direction of
higher surface tension, i.e., \emph{lower} surfactant
concentration along the air-water-swimmer contact line.
We note that this is opposite to the propulsion in the
direction of higher surfactant concentration for the
active liquid  swimmers discussed before.
The Marangoni boat mechanism is also employed by some
insects (rove beetle and  \emph{Velia}) \cite{Bush2005} to propel
on the water surface. 
There are also recent experiments \cite{Dietrich2020} and theoretical
work \cite{Wurger2014}  on a closely related system of thermally
driven Marangoni boats or surfers.

A full quantitative theory of Marangoni
boats including hydrodynamics,  surfactant
advection, direct Marangoni forces and
Marangoni flows  is still elusive despite previous
progress \cite{Soh2008,Nakata2015,Gidituri2019,JafariKang2020}.
Some theoretical approaches ignore
the advection \cite{Lauga2012,Wurger2014,Vandadi2017},
several ignore the hydrodynamic flow fields
\cite{Hayashima2001,Nagayama2004,Heisler2012,Iida2014,Suematsu2014,Koyano2017} 
or approximate it by uniform flow \cite{Boniface2019}, which clearly
oversimplifies the description of surfactant transport.
In particular,
  on the numerical side, a recent paper of Kang {\it et al.}
  \cite{JafariKang2020} provides progress by including  advection
  fully into the numerical solution for an
  anisotropic Marangoni boat.
A theoretical description
is complicated by the fact that most of the  Marangoni boats operate
at higher Reynolds numbers, and fluid flow generated during Marangoni
propulsion is typically featuring vortices \cite{Sur2019,JafariKang2020}.
Miniaturization  to the microscale leads to low Reynolds numbers.
Therefore, miniaturization is not only attractive for possible
applications but also provides a  starting point for the
development of hydrodynamic theories, as the  simpler
linear Stokes equation holds for fluid flow at low Reynolds numbers.
This has been initiated in Refs.\
\cite{Lauga2012,Wurger2014,Vandadi2017,Gidituri2019}.

Another question is regarding the role of intrinsic anisotropy,
namely, whether isotropic swimmers with no intrinsically defined
motion direction are also capable of a spontaneous motion which then
spontaneously breaks the symmetry of the system.
This question has been answered positively for autophoretic swimmers
\cite{michelin2013,Michelin2014}, where
it has been shown that advection by the surrounding fluid can maintain
the necessary  gradients in fields and/or concentrations
above a critical strength of the advection (characterized by a
dimensionless Peclet number). Liquid Marangoni swimmers
are always symmetric by construction
and have to maintain   an increased surfactant
concentration in the front  of their interface by
adsorption of surfactant (or micelles) or  by autophoretic
effects \cite{Yoshinaga2012,Izri2014,Schmitt2016}.
For the Marangoni boats the question regarding spontaneous symmetry breaking
has
been addressed experimentally in Ref.\ \cite{Boniface2019}, where
symmetric camphor disks have been shown to propel and
swimming velocities have been shown to be largely independent
of intrinsic swimmer anisotropy.
So far, a theoretical answer is missing for Marangoni boats.

\begin{figure}
  \centerline{\includegraphics[width=0.99\linewidth]{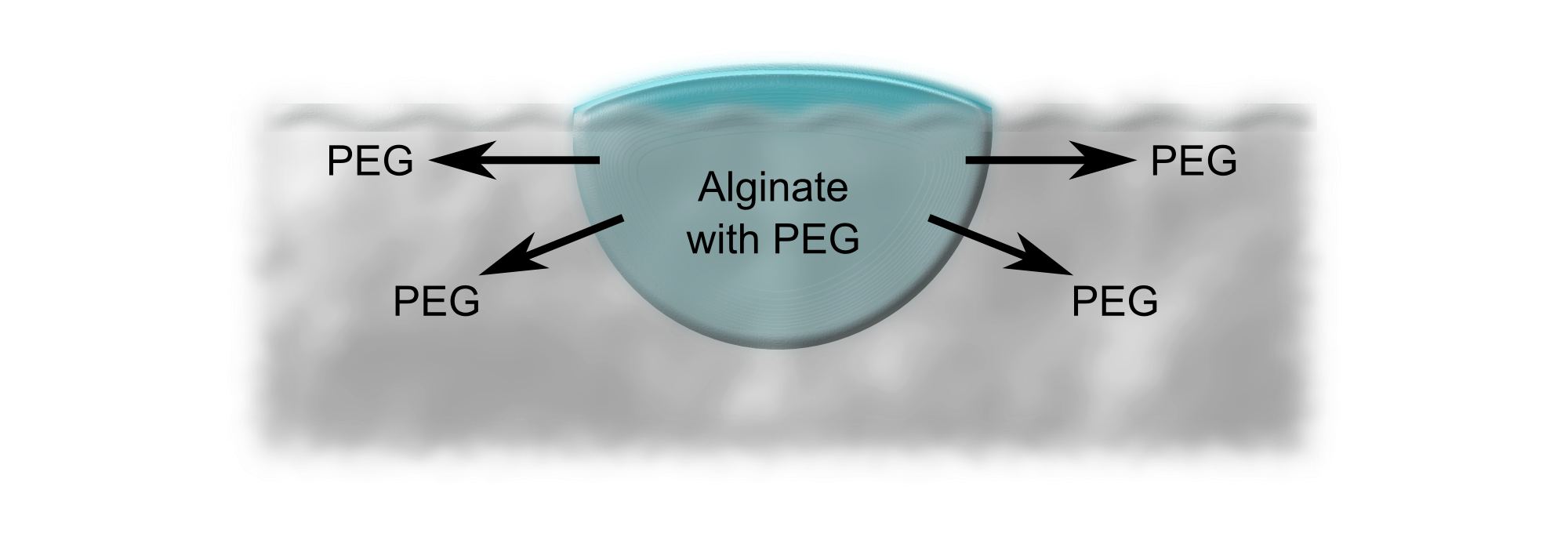}}
  \caption{Schematic  of the PEG-alginate capsule.
    The water-soluble ``fuel'' or spreading molecule PEG is
    incorporated during alginate capsule synthesis in the core and
   diffusively emitted during swimming.}
\label{fig:design}
\end{figure}

Here, we present a combined experimental and theoretical approach. 
We try to further approach the microscale by synthesizing 
alginate capsules as swimmer bodies, which provide
a porous matrix that can accept surface active molecules.
Several weakly surface active
fuels are tested, among which polyethylene glycol (PEG)
turns out to be the most effective.
The PEG-alginate swimmers exhibit fast and prolonged propulsion.
In general, we find prolonged propulsion only if 
spreading molecules are water-soluble as for PEG; then the 
air-water interface can regenerate by the fuel being dissolved
in water. 
The PEG swimmers are approximately half-spherical, i.e., symmetric;
therefore, we can address the question of spontaneous motion for a
 symmetric swimmer design. Moreover, a  half-spherical
 shape turns out to be very convenient for theoretical modeling,
 and has also been employed in Ref.\ \cite{JafariKang2020}.
For small Reynolds numbers, this geometry allows for a complete
theoretical description of Marangoni boat propulsion
 by analyzing 
  the coupled problems of surfactant
  diffusion and advection,  fluid flow, which includes 
  surfactant-induced fluid Marangoni flow, and surfactant adsorption
  at the air-water interface;
  we also include a possible evaporation of surfactant.
  The swimming speed is determined from the balance
  of Marangoni forces (both direct forces from surface tension gradients and
   from Marangoni flow forces) and drag forces. 
  We can address the problem of spontaneous symmetry breaking
  and predict the swimmer's speed in a stationary state. 
  This solution gives also hints how to generalize to
  higher Reynolds numbers using the concept of the Nusselt number,
  for which many results are known phenomenologically.

  On the experimental side, we find further effects, such as
  the repulsive  interaction of PEG-alginate swimmers
  with walls and the tendency to move in curved trajectories,
  which can be explained in the framework of  the Marangoni boat mechanism.

\section{Alginate based capsule swimmers}

\subsection{Swimmer synthesis and characterization}

The synthesized capsules show typical propelling mechanisms similar to
phenomena observed for the insect class of \emph{Microvelia}. Our artificial
microswimmers consist of PEG droplets, which were surrounded
by thin alginate shells (see Fig.\ \ref{fig:design}).
For the preparation of these particles, we first
form an aqueous PEG-alginate composite solution (standard:  $w_{\rm PEG
  300}=0.5\%$, $w_{\rm alginate}=0.5\%$). A droplet of this
mixture is  then deposed on the surface of an aqueous
$\ce{CaCl2}$ solution (standard: $w_{\ce{CaCl2}\cdot\ce{2H2O}}=0.5\%$).
The $\ce{Ca_{2+}}$
ions serve as cross-linker and induce, within several microseconds,
the gelation of the alginate membranes according to the
box-egg model \cite{Thiele1967,Leick2011,Klein1983,Cao2020}.
Immediately after the formation of these particles,
the capsules start to swim along the water surface.

\begin{figure}
  \begin{center}
     \includegraphics[width=0.99\linewidth]{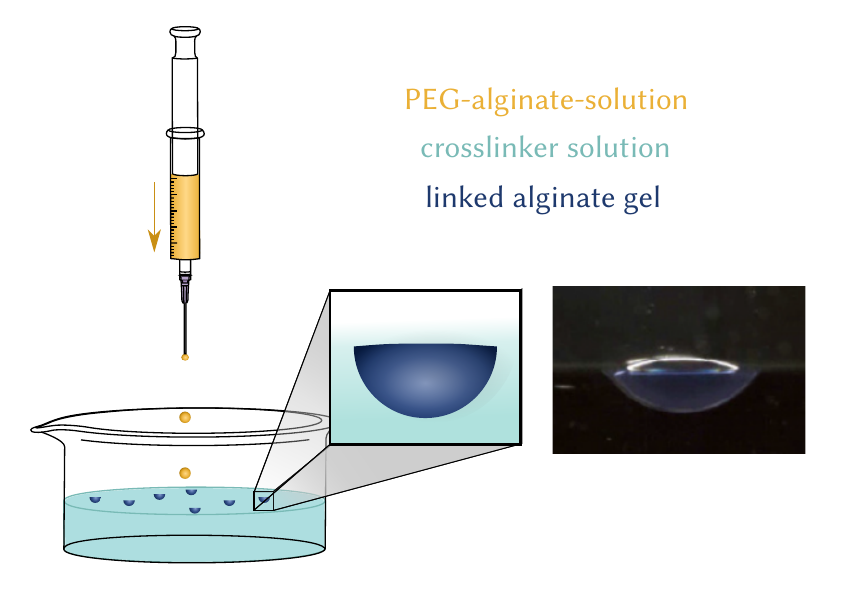}
  \caption{Synthesis of  PEG-alginate swimmers by pipetting
    microliter amounts of PEG-alginate solution into cross-linker solution.
   Side view of a   PEG-alginate swimmer showing its
     half-spherical shape.}
   \label{fig:dripping}
   \end{center}
\end{figure}

Dripping microliter amounts of
alginate into a cross-linker salt solution containing
counterions starts an ionotropic gelation  and 
produces approximately half-spherical alginate gel  capsules of millimeter
radius (see Fig.\ \ref{fig:dripping}) \cite{Thiele1967}.
We report results for 
$a\sim 1500\,{\rm \mu m}$; radii $a\sim 150\,{\rm \mu  m}$ can be reached.
For alginate gelation, different  salt solutions can be used containing
divalent  cations  such as 
$\ce{CaCl2}$, $\ce{CuCl2}$, or $\ce{BaCl2}$ solutions.

Adding surfactant to the alginate solution before dripping automatically 
loads the porous gel capsule with surfactant molecules.
For suitable surfactants,
capsules start to propel spontaneously on the air-water interface directly
after dripping.
The simple dripping technique allowed us to test
many different ``fuels'': successfully propelling fuels are
polyethylene glycols (PEGs) with molar weights $200-35000 \,{\rm g/mol}$,
alcohols,
acetic acid (stronger acids lead to protonization of alginate and subsequent
coagulation), and organic solvents.
A complete list of successfully tested fuels substances is given in
Table \ref{tab:fuels}.  Swimmers fueled by PEG (or  polypropylene glycol (PPG)),
in particular PEG 300, exhibit
the  best results regarding propulsion speed and
propulsion duration; the reason is a
suitable combination of diffusion constant, solubility, but also
gelation properties of the alginate-PEG mixture.
Corresponding monomers and dimers (ethylene glycol, propylene glycol,
diethylene glycol)
also exhibit good swimming properties but with lower speed and duration. 
It is particularly important  for a prolonged propulsion  that the
fuel substance lowers the air-water surface tension but
is also water-soluble such that it dissolves in the water reservoir
after spreading 
  in order to regenerate the air-water interface. 
  Evaporation from the air-water interface is another mechanism
  to achieve such a regeneration, which is at work in camphor
  boats \cite{Soh2008,Suematsu2014,Akella2018,Boniface2019}.
  Strong   surfactants and  detergents, such as sodium dodecyl sulfate,
  generate spreading pressures that can rupture the alginate capsule.
  Moreover, they quickly saturate the air-water interface such
  that concentration gradients and, thus, swimming can not be
  established.
In the following, 
we report results for solutions of 
alginate and PEG 300 ($w_{\rm alginate}=0.5\%$ and $w_{\rm PEG 300}=0.5\%$)
dripped into  a
$\ce{CaCl2}$ cross-linker
solution ($w_{\ce{CaCl2}\cdot\ce{2H2O}}=0.5\%$).

\begin{table*}
  \begin{center}
  \label{tab:fuels}
  \caption{Fuel substances leading to successful alginate capsule propulsion}
  \begin{tabular}{llll}
  \hline\noalign{\smallskip}
    Polymers   & Alcohols & Acids  & Organic solvents \\
    \noalign{\smallskip}\hline\noalign{\smallskip}
    PEG 200 & Ethylene glycol & Acetic acid &  Acetone \\
    PEG 300 & Propylene glycol &     &   Dimethyl sulfoxide\\
    PEG 400 &  Diethylene glycol &     &  Tetrahydrofuran \\
    PEG 600 &  Ethanol &             &   \\
    PEG 1000 &  Isopropanol  & &\\
    PEG 6000 & 1-pentanol & &\\
    PEG 20000 & Benzyl alcohol & &  \\
    PEG 35000 & 1-hexanol  & &\\
    PPG 400  & 2-butanol, Tert-butanol & & \\
               & Dodecanol & &\\
   \noalign{\smallskip}\hline
  \end{tabular}
  \end{center}
 \end{table*}

\begin{figure}
  \centerline{\includegraphics[width=0.99\linewidth]{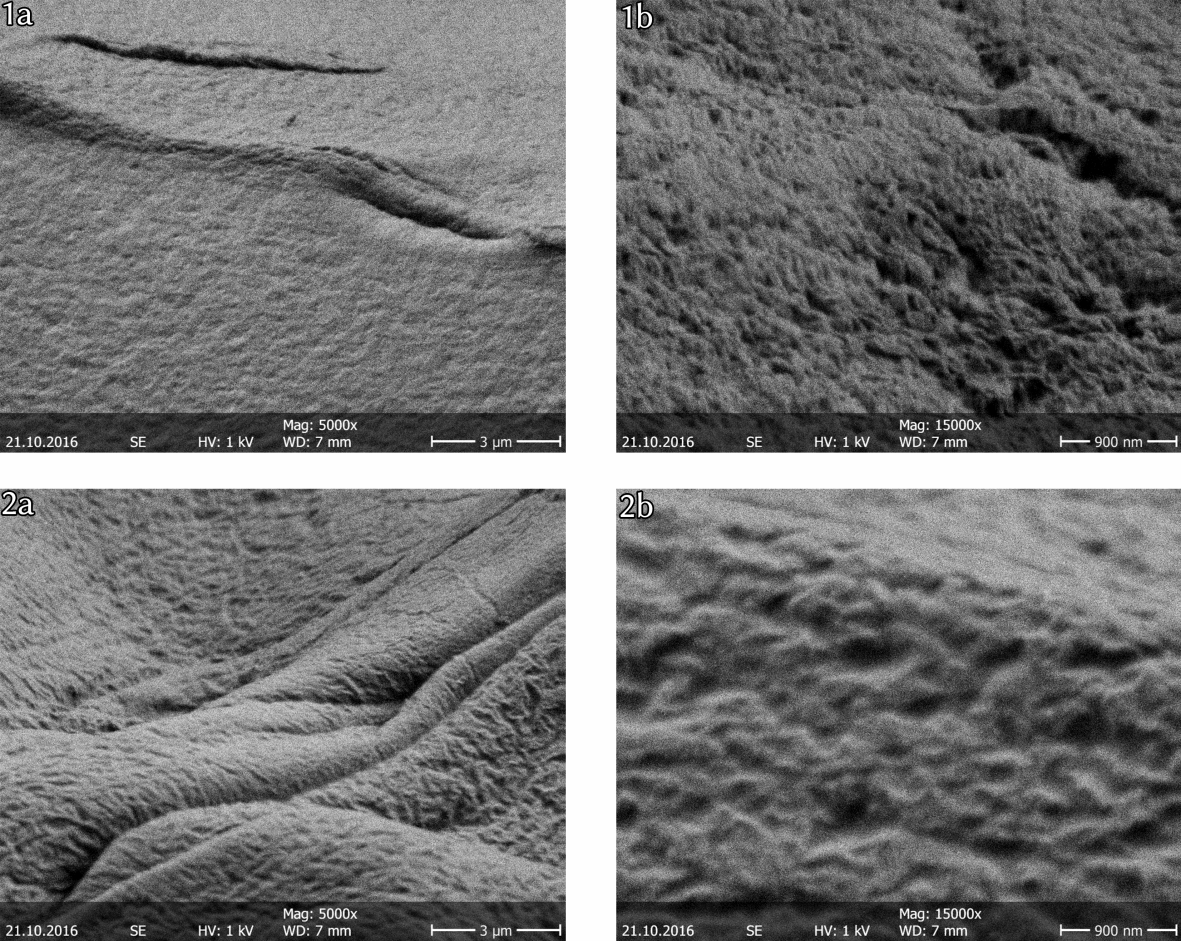}}
  \caption{Scanning electron microscopy images of the porous
    structure of unloaded alginate capsules (1a,1b) and PEG-loaded
    alginate capsules (2a,2b) in 5000-fold (1a,2a) and 15000-fold (1b,2b)
      magnification.}
\label{fig:SEM}
\end{figure}

Alginate gels have a porous structure \cite{Thiele1967,Leick2011,Klein1983}.
Scanning electron microscopy (SEM) of the alginate capsules
reveals their porosity and also a certain 
roughness on the microscale with asperities on the capsule
surface (see Fig.\ \ref{fig:SEM}).
The pores are essential for the slow diffusive emission of surfactant
from the capsule \cite{Leick2011,Klein1983}.
PEG diffusion through the porous alginate matrix is much
slower than PEG diffusion in water; therefore, PEG should be  released
with a slowly varying  controlled diffusive current that  is 
limited by  its slow diffusion in the alginate.  
The shape of the capsule and the spatial distribution
of pores on the surface can break the overall spherical symmetry and
give rise to small anisotropies in the emission, in principle.

\subsection{Swimming motion}

\begin{figure}
\centerline{\includegraphics[width=0.99\linewidth]{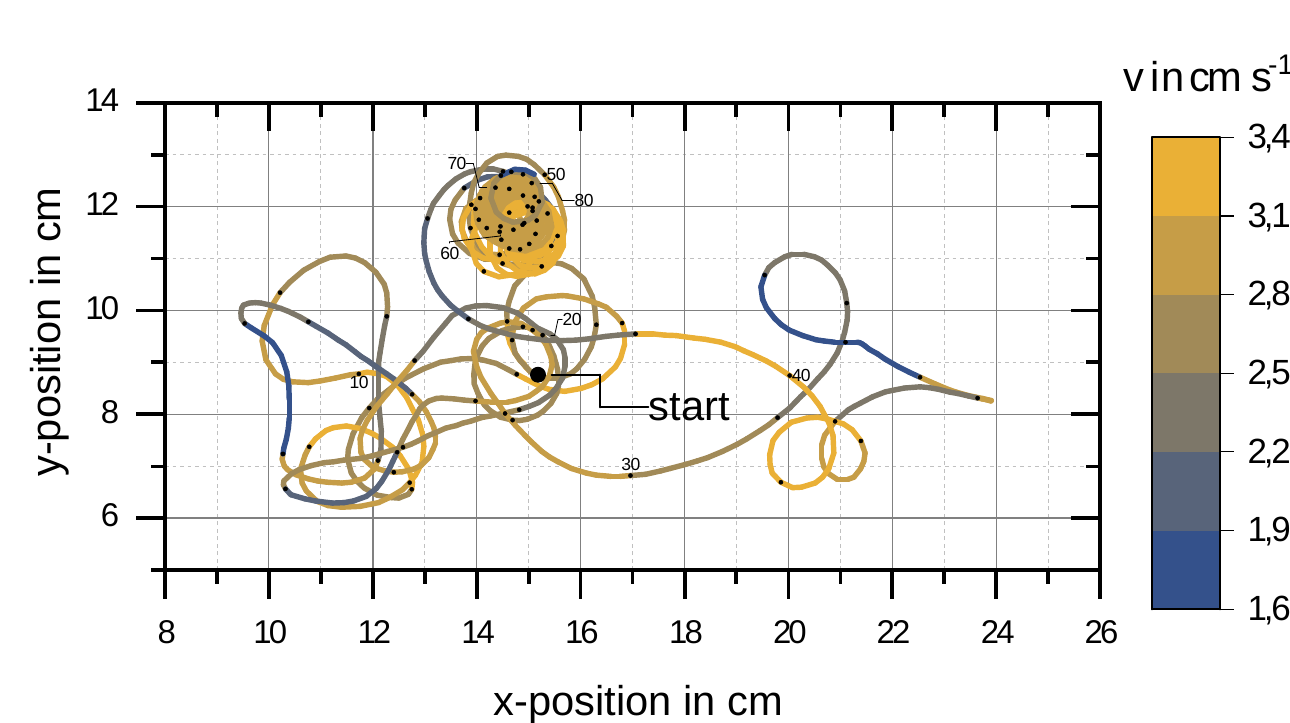}}
  \caption{Typical swimming trajectory of a   PEG-alginate swimmer
    in the cross-linker solution. Color-coded is the swimming velocity
    $U_{\rm swim}$,
     the trajectory shows the first  $84\,{\rm s}$ of swimming.}
\label{fig:trajectory}
\end{figure}

The alginate-PEG swimmers exhibit a 
fast and sustained motion.
The swimming motion was observed in a cylindrical dish
(diameter $24\,{\rm cm}$) for up to $20\,{\rm min}$. 
The swimmers exhibit typical speeds
  $U_{\rm swim} \sim
2-3\, {\rm cm/s}$  corresponding to $10-20$ swimmer sizes per second
(see Fig.\ \ref{fig:trajectory}); after $20\,{\rm min}$,
velocities $U_{\rm swim} \sim 1 \,{\rm cm/s}$ can still be measured. 
This swimming performance is comparable to camphor boats
\cite{Akella2018,Boniface2019} and active liquid droplets
\cite{Izri2014,Herminghaus2014}.

A typical swimming trajectory (lasting $84 \,{\rm s}$) far from a wall
is shown in Fig.\
\ref{fig:trajectory}. We obtained this
trajectory from a single particle tracking analysis (using ImageJ);
typical swimming velocities are $U_{\rm swim} \sim
2-3 \,{\rm cm/s}$  corresponding to $10-20$ swimmer sizes per second.
This corresponds to moderate Reynolds numbers 
${\rm Re} = {\rho U_{\rm swim} 2a}/{\mu}\sim 60$  
(with the swimmer diameter $2a\simeq 3000\,{\rm \mu m}$ as length scale
and 
the viscosity and density of water, $\mu \simeq 10^{-3} \,{\rm Pa s}$ and 
 $\rho = 10^3 \,{\rm kg/m^3}$).  

 \begin{figure}
   \begin{center}
     \includegraphics[width=0.99\linewidth]{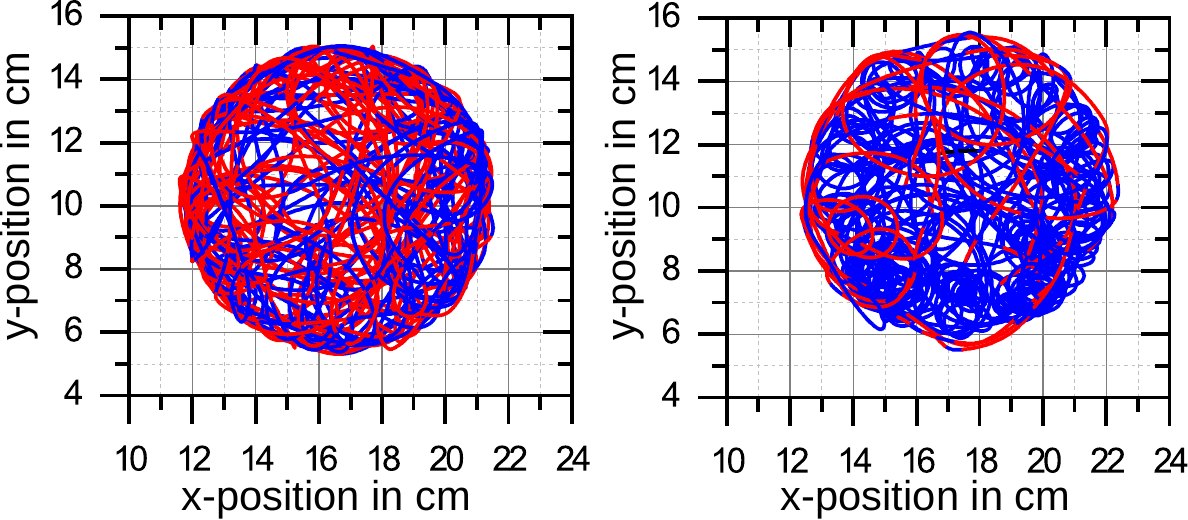}
  \caption{Swimming trajectory of two  PEG-alginate swimmers
    in a cylindrical container. Color-coded is the  sign of the curvature,
    blue/red trajectories curve clockwise/counter-clockwise.
    Swimmers prepared according to same protocol exhibit individually
    different curving behavior (left: mostly counter-clockwise, right: mostly
    clockwise). Reflections
    at walls are of different duration. 
     }
     \label{fig:trajectory2}
 \end{center}
\end{figure}

Swimming trajectories such as in  Fig.\
\ref{fig:trajectory} and in confinement in Fig.\
\ref{fig:trajectory2}  exhibit phases with a characteristic curvature.
Marking the swimmer with elongated plastic fragments shows that
the elongated fragment is not turning with respect to the direction of motion,
 i.e., the curving of the trajectory is correlated with a reorientation of the
 swimmer.  This is a hint 
 that during curved swimming the  swimming direction is
 linked to the orientation of the particle and, therefore,
 that the spherical symmetry is slightly broken by irregularities in
 the porous structure
 of the alginate particle (see SEM pictures in Fig.\ \ref{fig:SEM}).
 The swimming direction is selected by
 dominating pores which  determine  a preferred direction of
 emission and, thus, propulsion by the resulting surfactant
 gradients. 
 Curving itself can  be  caused by additional torques
 from  asperities of the alginate capsule where surfactant
 is emitted preferentially in the tangential direction.
 A similar mechanism is at work at camphor-driven rotors
 \cite{Nakata1997,Koyano2017}.
 This is supported by the finding that the curving behavior of
 swimmers prepared by the same protocol (such as the swimmers in
 Fig.\ \ref{fig:trajectory2}) is individually different and seems
 to depend on small differences between irregularities
 acquired in the preparation process.  Recently,
 also vortex shedding at Reynolds numbers ${\rm Re}\sim 100-200$ 
 have been proposed to cause curving  of trajectories \cite{Sur2019}.

Swimmers are also repelled by walls and reverse their direction of motion
normal to the wall. In a course of a   collision in normal direction,
the swimmer keeps, however, its orientation while the
direction of motion is reversed, i.e., during normal wall collisions the
 swimming direction also reverses with respect to the
  particle orientation.
Swimming direction reversal has also been observed  for camphor
boats \cite{Hayashima2001,Nagayama2004,Nakata2015}.
Swimming trajectories  in  Fig.\
\ref{fig:trajectory2} also show collisions with walls that  last longer;
these collisions can also feature a reorientation of swimmer,
similar to what has been  predicted for self-diffusiophoretic swimmers
\cite{Uspal2015}.

\subsection{Swimming mechanism}

The order of magnitude of swimming speeds can only be explained
as a result of a modulation of the large liquid-air surface tension.
Marangoni mechanisms based on surface tension variations
within the gel-liquid interface between alginate capsule and surrounding water
are unlikely because the interfacial tensions and, thus,
also Marangoni stresses, are too small for solid-liquid or gel-liquid
interfaces.
This hints at a Marangoni boat propulsion mechanism for the alginate-PEG
swimmers.

There is further 
experimental evidence supporting the Marangoni boat mechanism:
(a) Sinking capsules stop swimming which excludes a phoretic or
Marangoni mechanism
based only on the swimmer-liquid interface such as
the  active liquid droplet mechanisms
\cite{Izri2014,Herminghaus2014}.
(b) Only water-soluble spreading molecules lead to prolonged propulsion
because they allow regeneration of the air-water interface by
re-dissolving after spreading, which is crucial to establish
concentration gradients at the air-water interface.
(c) Local Wilhelmy plate surface tension measurements demonstrate
 surface tension modulations
 depending on the  distance to a swimmer; this demonstrates
 that additional surfactant is emitted close to the swimmer. 
 (d) Particle image velocimetry (PIV)
 measurements and selective staining with aniline
 show fluid motion consistent with surfactant
 spreading by surface tension reduction.
(e)
Swimming speed depends on the  diffusive mass outflux. We will
develop a theory for this dependence in the theoretical part
of the paper, which describes the data (without free fitting parameters). 
(f)
Repulsion and direction reversal without reorientation of the swimmer
can be explained by an accumulation of surfactant emitted by the
swimmer in front of the wall.
This points to a motion opposite to  the surfactant concentration gradient,
while the  direction of motion is not completely
fixed relative to swimmer orientation.

More details regarding points (d) and (e) are given below.
  All these results suggest that the swimmer diffusively emits surfactant
  which reduces the surface tension. The swimmers are spherically
  symmetric to a good approximation and this symmetry is strongly broken
  by the concentration profile
  in the fast moving state. The only available mechanism for 
  symmetry breaking in the moving state is by advection
  to the surrounding moving liquid, which selects a swimming
  direction spontaneously.

 The results regarding curved trajectories and wall collision suggest
 that spherical symmetry is not perfect and large pores in the capsule
 shell can
 select a weakly preferred propulsion direction and link capsule
 orientation to swimming direction. This weak link can be
 deleted during a normal collision with a wall, when the swimmer
 reverses direction without changing orientation.

\subsubsection{PIV measurements}

\begin{figure*}
  \begin{center}
     \includegraphics[width=0.99\linewidth]{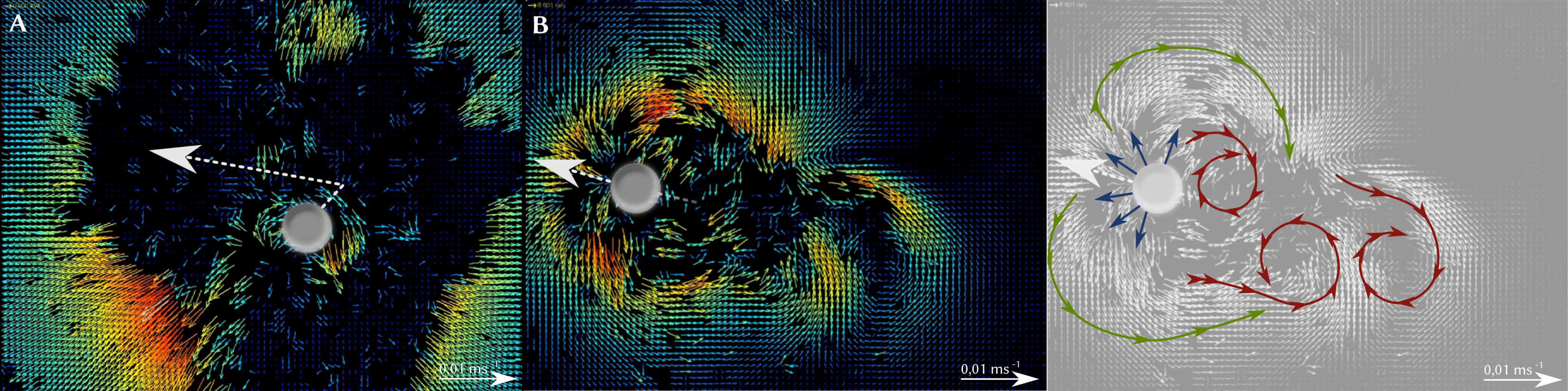}
    \caption{PIV measurements of a PEG-alginate swimmer (A) after
      dripping and (B) in motion. The velocity scale bar refers to the
      fluid velocities, which are also color-coded for velocity.
      The white dashed arrows indicate the direction of the particle
      motion. On the right we indicate the fluid motion in (B)
      featuring radial Marangoni flow (blue), combined with flows
       corresponding to two counter-rotating vortices 
        created by particle motion (green),
      and downstream Karman-like vortices (red). }
    \label{fig:PIV}
    \end{center}
\end{figure*}

PIV measurements were performed with polymethyl methacrylate (PMMA)
tracer particles with sizes between $30$ and $50\,{\rm nm}$ and visualize the
fluid flow close to the air-water interface. Figure \ref{fig:PIV}
shows the results directly after swimmer synthesis by dripping (A),
i.e., in the initial starting phase of the swimmer and (B) shortly
after the swimming started.

In Fig.\ \ref{fig:PIV}(A) we observe strong radial spreading of
surfactant by initial Marangoni flows. Then, the symmetry is
spontaneously broken when swimming is initiated and
Fig.\ \ref{fig:PIV}(B) shows the
fluid surface flow in the initial swimming stage.
During swimming we still observe radial Marangoni flows (blue) but
the fluid flow around the swimming object creates a tangential
backward component (green) because two counter-rotating
vortices form; moreover, Karman-like vortices
appear on the rear side (red). Vortex formation demonstrates that the
fluid motion happens at moderate  Reynolds numbers ${\rm Re} \sim 60$.
Similar vortex structures have also been observed in Ref.\ \cite{Sur2019}
for disks propelled by alcohol.
  Nevertheless, Reynolds numbers are moderate (${\rm Re} \ll 200$)
  such that we can expect a steady fluid flow 
  (eventually with boundary layer separation from the sphere
  and stationary
Marangoni vortices). Only
at higher Reynolds numbers ${\rm Re} > 200$,
we expect unsteady or even turbulent flow around a sphere \cite{Johnson1999}.

\subsubsection{Mass outflux and velocity measurements}

We propose that 
the swimming motion is caused by surfactant that is diffusively
emitted by the PEG-alginate capsule.
Therefore, a relative slow
reduction of the total mass of the PEG-alginate capsules should
be measured, which also correlates with the swimming speed.
 Overall spherical symmetry
of the capsule implies that the 
emission current density $\alpha$ is uniform on the capsule
surface.

\begin{figure*}
  \begin{center}
     \includegraphics[width=0.99\linewidth]{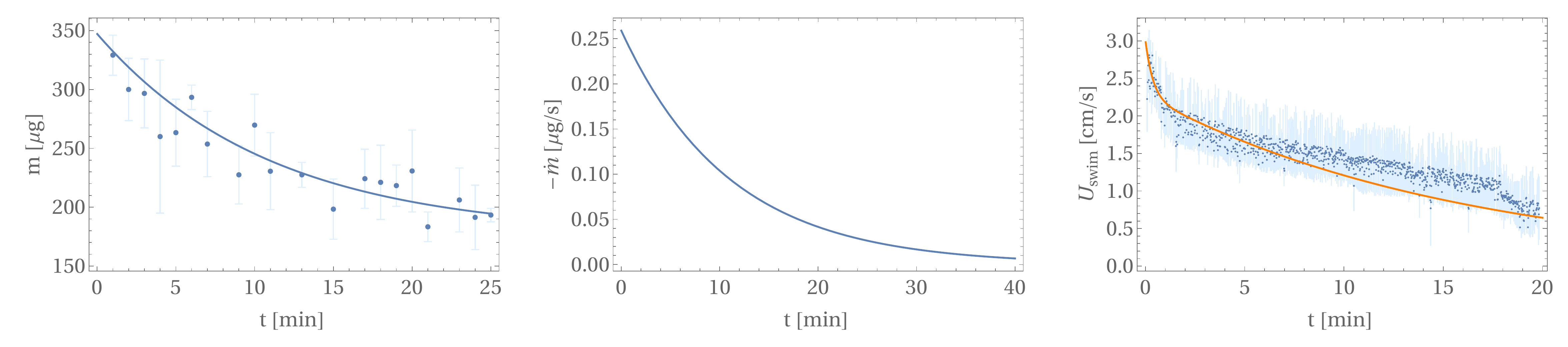}
  \caption{Left: Mass of the PEG-alginate swimmers as a function of time
   with  an exponential fit $m(t) = m_\infty + m_0\exp(-t/\tau_m)$ (see text);
   error bars (light blue) denote the
     standard deviation.
    Middle: Resulting mass loss $-\dot{m}$ as a function of time.
    Right: Corresponding velocity of the swimmer averaged
     over 10 swimmers; error bars (light blue) denote the
     standard deviation. The orange line is a fit
     $U_{\rm swim}(t) =
     u_0\exp(-t/\tau_{u,0})+ u_1\exp(-t/\tau_{u,1}) $ (see text)
      motivated by  the existence of two swimming phases. }
   \label{fig:mdot}
   \end{center}
\end{figure*}

Quantitative measurements of the mass outflux are difficult.
In Ref.\ \cite{Boniface2019} this has been achieved only indirectly
by measuring the increase in surfactant in the surrounding solution.
Here, we measure the mass outflux directly by removing swimmers
(prepared according to the same protocol) after times
$t=1,2,3,... {\rm min}$ from the swimming solution, dry freezing
the swimmers to completely remove water from the alginate hydrogel,
and determine their weight, which  gives the mass $m(t)$ of the swimmer
at times $t=1,2,3,... {\rm min}$.
Figure \ref{fig:mdot}(left) shows the results for the mass
 averaged over 10 swimmers,
 error bars are the standard deviation.
 
Diffusive outflux through a porous shell 
of thickness $h$ ($h<a$) approximately follows  an exponential
decay. The emission current density
is $\alpha \approx    -D_s (c_i-c_0)/h$,
where $D_s$ is the diffusion constant in the gel, $c_i$ the interior
PEG concentration and $c_0$ the exterior PEG concentration in solution.
We assume that PEG has to diffuse over a fixed distance $h$
for release; more refined  release models use a time-dependent
diffusion distance \cite{Higuchi1961,Higuchi1963,Boniface2019}.
For a half-sphere, this results in
$\dot{c}_i = 3\alpha/a =  -3 D_s (c_i-c_0)/ha$
and an exponential decay of $c_i(t)$ and, thus, $m(t)$.
This motivates an error-weighted  least-square fit with an  exponential
\begin{equation}
  m(t) = m_\infty + m_0\exp(-t/\tau_m),
\end{equation}
which describes the data well with an ``empty'' mass
$m_\infty \approx 180\,{\rm \mu g}$,  a mass loss
$m_0\approx 182\, {\rm mu g}$,
and a time constant $\tau_m \approx 9.95\,{\rm min}$
(see Fig.\ \ref{fig:mdot}(left)).
This confirms a slow  diffusional surfactant release, i.e., $\alpha$
is changing on a large time scale $\tau_m$;
this time scale  is large compared to
any microscopic time scale of the fluid flow and the surfactant diffusion. 
Therefore, we  expect that fluid flow
and surfactant concentration are always in a quasi-stationary state
during swimming, i.e., adiabatically following a slowly changing $\alpha$.
Differentiating with respect to time gives the
mass outflux $\dot{m}$ as a function of time
[see Fig.\ \ref{fig:mdot}(middle)].

The corresponding swimming velocity $U_{\rm swim}$
of the PEG-alginate swimmers is measured via the single
particle tracking analysis.
Figure \ref{fig:mdot}(right) shows the results for the velocity
 averaged over 10 swimmers
 prepared according to the same protocol as for the mass measurements,
 error bars are the standard deviation. 
 The data clearly shows a fast initial drop of the velocity in a first phase,
 followed by a slower decay in a second phase.
 During the first phase, slow diffusion of surfactant through
 the porous alginate matrix might not be necessary yet and gelation
 of the capsule alginate shell might also be incomplete.
 The existence of several swimming phases has also been
 observed for  camphor disks in Ref.\ \cite{Akella2018}.
 The second phase should be characteristic for propulsion triggered
 by slow diffusional release as described above.
 This motivates a fit
 \begin{equation}
   U_{\rm swim}(t) =
   u_0\exp(-t/\tau_{u,0})+ u_1\exp(-t/\tau_{u,1})
 \end{equation}
 with  two
 exponentials. The resulting error-weighted least-square fit
 describes the data well as shown in Fig.\  \ref{fig:mdot}(right).
 The first phase has a time constant $\tau_{u,0} \simeq 0.37\,{\rm min}$
 (and $u_0\simeq 0.74\,{\rm cm/s}$),
 whereas the second phase has  $\tau_{u,1} \simeq 15.89\,{\rm min}$
 (and $u_1\simeq 2.27\,{\rm cm/s}$), which is comparable with $\tau_m$.
 This further supports  that diffusive release of surfactant causes
  the propulsion during the second phase.

\section{Theoretical results}

\subsection{Model}

In the theoretical part of the paper, we focus on the 
 dependence of swimming speed on diffusive
 surfactant release. So far, this important
   question has not received attention
   in the literature.
 The strategy to calculate
 the swimming speed is as follows. 
 We first prescribe a stationary
velocity $\vec{U} = U \vec{e}_z$
of the swimmer and  analyze
the following three coupled problems for their stationary state:
\begin{itemize}
  \item[(i)] Surface tension reduction by surfactant adsorption
  at the air-water interface;
  depending on the volatility of the surfactant 
  we also need to include a possible evaporation of surfactant.
 PEG is not volatile but the theory will apply to the physics of
  the Marangoni boat mechanism in general and should also explain
  quantitative  results on camphor boats from Ref.\ \cite{Boniface2019}.
  As opposed to PEG, camphor is a volatile surfactant which quickly
  evaporates from the air-water interface. 
\item[(ii)]
   Fluid flow, which includes 
   both the fluid flow induced by motion of the half-spherical
   capsule and additional surfactant-induced  Marangoni flow inside
     the fluid.
\item[(iii)]  
  Diffusive surfactant release from the swimmer and subsequent 
  diffusion and advection.
\end{itemize}
Solving these three coupled problems we can obtain the
Marangoni forces as a function  of the prescribed velocity $U$
from the surfactant concentration profile. 
Finally, the actual swimming velocity $U=U_{\rm swim}$
 is determined by the force equilibrium between  drag force,
  direct propelling Marangoni forces from the surface tension gradient along the
 air-water-swimmer   contact line, and  Marangoni   flow forces,
  which can increase either the drag or the direct  Marangoni propulsion
  force.

The fluid flow part (ii), the drag force,
and also the Marangoni forces in the force balance
strongly depend on the Reynolds number. 
Although the Reynolds number for the PEG-alginate swimmers is moderate
 (${\rm Re} \sim 60$),
  we will first develop a low Reynolds number theory, and try
  to address higher  Reynolds numbers afterward using
   phenomenological results for the Nusselt number.

\begin{figure}
  \centerline{\includegraphics[width=0.99\linewidth]{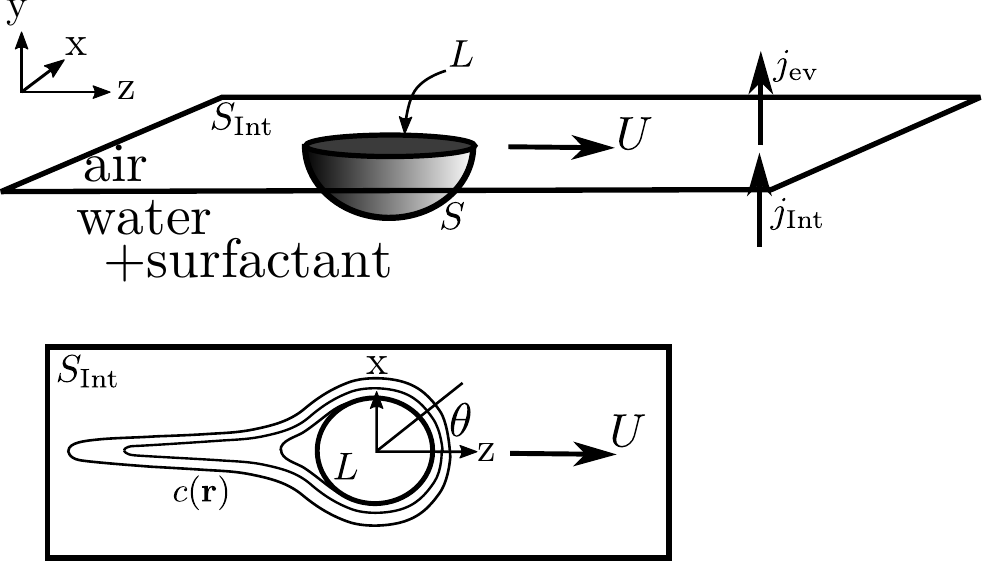}}
  \caption{Side view (top) and top view (bottom) of
    the  half-spherical Marangoni swimmer geometry with surfactant
     concentration field $c(\vec{r})$ and coordinates. 
     }
\label{fig:scheme}
\end{figure}

   We introduce coordinates such that the origin $r=0$
is at the center of the circular
planar solid surface of the half-sphere, and  the liquid-air interface is at
   $y=0$ (with $y<0$ being the liquid phase) and
   $\vec{e}_z$ will coincide with
the spontaneously selected swimming direction.
We also use spherical coordinates such that $\theta = 0$ is the swimming
direction and
the interfacial plane is located at $\phi = 0,\pi$ ($y=0$).
The half-sphere has  radius $a$ such that the  contact line
is at $r=a$  and $\phi = 0,\pi$ (and parametrized by $\theta$).
We denote the half-spherical surface of the swimmer by $S$,  the
circular  air-water-swimmer  contact line by $L$,
and the liquid-air interface
outside the swimmer  as $S_{\rm Int}$, see Fig.\ \ref{fig:scheme}.

\subsubsection{Surface tension  reduction by surfactant adsorption and evaporation}

We start with problem (i), which is independent of the
Reynolds number.  In equilibrium,
the  surfactant concentration $\Gamma(\vec{r})$
at the interface $y=0$ for a given bulk subsurface 
concentration $c(\vec{r})$ is given by  Langmuir  adsorption 
\begin{equation}
  \Gamma(c) = \Gamma_{\rm max} {K_Lc}/(1+K_Lc)
 \end{equation}
(with the adsorption equilibrium constant  $K_L$
and  the maximal surfactant surface
concentration $\Gamma_{\rm max}$).
In Langmuir adsorption,
we assume ideal behavior of the surfactant  molecules. 
According to the Gibbs adsorption isotherm,
the   interfacial  surface tension $\gamma$
is related by $d\gamma = - RT \Gamma d\ln c$
to  surface concentration and  bulk concentration \cite{Rosen2012}.
Together with the Langmuir equation, this leads to the Szyszkowski equation
\begin{align}
  \Delta \gamma  &= -RT\Gamma(c_0) \frac{1}{c_0} \Delta c
            =       -R T \Gamma_{\rm max}\frac{ K_L}{1+K_Lc_0}\Delta c
  \label{eq:Szyszkowski}
\end{align}
(with the gas constant  $R=N_A k_B$ and $\Gamma_{\rm max}$ in mol per area),
formulated for 
 small local variations around a background, $c(\vec{r}) =
 c_0 + \Delta c(\vec{r})$.
 Small surfactant concentration variations
 thus
linearly reduce the surface tension, 
\begin{align}
  \Delta \gamma(\vec{r})  &= -\kappa \Delta c(\vec{r})
                          ~~\mbox{with}
~~  \kappa = RT \Gamma_{\rm max} \frac{K_L}{1+K_Lc_0},
  \label{eq:kappa}
\end{align}
where $\vec{r}$ is an interfacial  vector with $y=0$.
 We choose the  background $c_0$  as the bulk value $c_0 = c(\infty)$ 
for $|\vec{r}|\to \infty$.

In formulating Eq.\ (\ref{eq:kappa}) locally,
we already  assumed that the on and off kinetics
of surfactant to the interface is fast
such that equilibrium can be assumed
to be established instantaneously at \emph{every} point $\vec{r}$
on the interface.
Then, also the surface concentration $\Gamma(\vec{r})$ is slaved to the bulk
and only a passive  ``reporter''
of the bulk subsurface concentration $\left.c(\vec{r})\right|_{y=0}$,
and  we do not have to solve 
a separate dynamics for $\Gamma(\vec{r})$ in the interface.
This assumption is typically valid for small surfactant molecules
\cite{Li1994}, in particular for water-soluble spreading molecules
such as PEG.
The assumption also implies that there is no flux imbalance within the
interface, and also the bulk diffusive flux to the
interface $S_{\rm Int}$ has to vanish, 
\begin{equation}
  j_{\rm Int} =
  -\left. D \vec{\nabla} c(\vec{r})\cdot \vec{n}^{\rm out}\right|_{y=0}
  =-D \left. \partial_y  c(\vec{r})\right|_{y=0}=   0,
  \label{eq:jInt}
\end{equation}
which provides the corresponding boundary condition to the
diffusion-advection problem (iv) in the bulk.
Here, $D$ is the surfactant diffusion constant in the bulk liquid. 
The surface concentration  $\Gamma(\vec{r})$ should also be small
enough to avoid saturation of the air-water interface, which
also requires  water-soluble spreading molecules
such as PEG.

So far, we did not consider the possibility of surfactant evaporation
from the interface. This enters the balance of fluxes to and from the
interface (see Fig.\ \ref{fig:scheme}),
and  we have to replace Eq.\ (\ref{eq:jInt}) by
\begin{equation}
  j_{\rm Int} =
  -\left. D \vec{\nabla} c(\vec{r})\cdot \vec{n}^{\rm out}\right|_{y=0}
  = -  j_{\rm ev} =  k\left. c(\vec{r})\right|_{y=0},
  \label{eq:jev}
\end{equation}
where $k$ is the rate constant for evaporation.

\subsubsection{Fluid flow at low Reynolds numbers}

We consider the rest frame of the  swimmer 
and linearly decompose the total fluid flow field into
a field $ \vec{v}(\vec{r})$, which is the flow field of a
half-sphere pulled with velocity $U\vec{e}_z$ through the liquid
and a correction $\vec{{v}}_{\rm M}(\vec{r})$ from Marangoni flows, 
$ \vec{v}_{\rm tot}(\vec{r})
              = \vec{v}(\vec{r}) +
              \vec{{v}}_{\rm M}(\vec{r})$.
For low Reynolds numbers, \emph{both} $ \vec{v}(\vec{r})$ and $\vec{{v}}_{\rm
  M}(\vec{r})$ (and the associated pressure fields)
fulfill the incompressibility condition $\vec{\nabla}\cdot \vec{v} = 0$ and
the linear Stokes equation  $\mu \vec{\nabla}^2 \vec{v}
= \vec{\nabla} p$, where $\mu$ is the fluid viscosity.
The Stokes equations for $\vec{v}(\vec{r})$ and
$\vec{v}_{\rm M}(\vec{r})$ are decoupled because of linearity;
this will be  different at high Reynolds numbers.

The flow field  $\vec{v}(\vec{r})$ 
of an externally pulled half-sphere has no-slip boundary
conditions on the surface of the sphere, 
stress-free boundary conditions at the liquid-air interface,
and $\vec{v}(\infty) =  - U\vec{e}_z$ at infinity.
The total flow field  $\vec{v}_{\rm tot}(\vec{r})$ also  has
 no-slip boundary
 conditions on the surface of the sphere,
 assumes $\vec{v}_{\rm tot}(\infty) =  - U\vec{e}_z$ at infinity,
 but is subject to Marangoni stresses
 at the liquid-air interface.
 Consequently, the difference
 $\vec{v}_{\rm M}(\vec{r}) =  \vec{v}_{\rm tot}(\vec{r})-
\vec{v}(\vec{r}) $ from Marangoni flows has no-slip boundary
 conditions on the surface of the sphere, has vanishing velocity 
 $\vec{v}_{\rm M}(\infty) = 0$ at infinity, and
 is subject  to Marangoni stresses
 at the liquid-air interface.
Moreover, for all three flow fields, there is no normal flow across the
liquid-air interface $\left. v_{\rm tot, y}(\vec{r}) \right|_{y=0} =
\left. v_y(\vec{r}) \right|_{y=0} =  \left. v_{\rm M, y}(\vec{r})
\right|_{y=0} = 0$. 
We will assume that the liquid-air interface remains flat,
even if the sphere moves. This requires that typical viscous forces
remain small compared to interfacial stress,
$\mu U  \ll \gamma$, which is fulfilled with
$\mu U \sim  10^{-5}\, {\rm N/m}$ for
$U\sim 1 \,{\rm cm/s}$ and $\gamma \sim  0.07 \,{\rm N/m}$ for the air-water
interface. We also neglect a possible  curvature of the interface from wetting
effects. 

The Marangoni flow is caused by  tangential Marangoni
stresses at the liquid-air interface $y=0$,
\begin{equation}
  \mu  \vec{n}^{\rm out}\cdot \vec{\nabla}
  \left. \vec{v}_{\rm M}(\vec{r})\right|_{y=0} =
  \mu \partial_y  \left. \vec{v}_{\rm M}(\vec{r})\right|_{y=0}
  = \vec{\nabla}_S \Delta \gamma(\vec{r}),
  \label{eq:Marangoni}
\end{equation}
which act both on $ \vec{v}_{\rm M}$ and $ \vec{v}_{\rm tot}$.

At low Reynolds numbers, the flow field $\vec{v}(\vec{r})$
is given by ``half'' ($y<0$)
the Stokes flow field around a sphere, which automatically
fulfills the boundary condition  $\left. v_y(\vec{r}) \right|_{y=0} = 0$
for symmetry reasons. In spherical coordinates, the axisymmetric
Stokes flow field is 
\begin{subequations}
\begin{align}
       \vec{v}(\vec{r}) &=  U \cos\theta  u(r/a)\vec{e}_r +
                          U \sin\theta  v(r/a) \vec{e}_\theta
                    ~~\mbox{with} \nonumber\\
  u(r/a) &= \left[-\frac{1}{2} \left(\frac{a}{r}\right)^3 +
                \frac{3}{2} \frac{a}{r} - 1 \right]
         \label{u}\\
  v(r/a)
    &= \left[-\frac{1}{4} \left(\frac{a}{r}\right)^3 -
               \frac{3}{4} \frac{a}{r} + 1 \right]
               \label{v}
\end{align}
\end{subequations}

\subsubsection{Surfactant diffusion and advection}

Surfactant molecules are emitted from the
half-spherical surface $S$ and diffuse into the liquid phase.
At the same time, they are advected by the total fluid flow. 
In the stationary limit, the bulk concentration field
is governed by the diffusion-advection equation
\begin{align}
  0=  \partial_t c &= D \vec{\nabla}^2 c -
      (\vec{v}(\vec{r})+ \vec{v}_{\rm M}(\vec{r})) \cdot \vec{\nabla} c.
          \label{eq:diffadv0}
\end{align}  
Because of the  slow  diffusional surfactant release
the appropriate boundary condition on $S$ is a constant
flux boundary condition,
\begin{align}
   \left. \vec{j}\cdot\vec{n}\right|_S
  &=- D\left. \vec{\nabla} c\cdot\vec{n} \right|_S  =\alpha,
    \label{eq:constantflux}
\end{align}
together with $c(\infty)=0$ and  the no-flux
boundary condition (\ref{eq:jInt}) at the
interface $S_{\rm Int}$.
The flux $\alpha$ is only slowly changing  (on the time scale $\tau_m$)
and approximated as a constant for the calculation of quasi-stationary
fluid flow and
concentration fields.

\subsubsection{Drag and Marangoni forces at low Reynolds numbers}

The half-spherical swimmer moving at velocity $U$
is subject to three forces. First, there is
the drag force, which is, at low Reynolds numbers, given by
the Stokes drag for a half-sphere,
\begin{equation}
   \vec{F}_{\rm D} = F_{\rm D}\vec{e}_z =
     -3\pi \mu a U \vec{e}_z.
   \label{FD}
 \end{equation}
 Second, there is the direct Marangoni propulsion
 force  $\vec{F}_{\rm M} = F_{\rm M} \vec{e}_z$ from
 integrating the  surface stress $\Delta\gamma(\vec{r})= -\kappa c(\vec{r})$
along the air-water-swimmer  contact line,
\begin{align}
  F_{\rm M} &\equiv 
              -\kappa  \oint_L ds  (\vec{e}_n\cdot \vec{e}_z) c(\vec{r})
           \nonumber\\
  &= - 2 \kappa a \int_0^\pi d\theta \cos\theta c(a,\theta)|_{y=0}.
  \label{FM}
\end{align}

Third, there is the Marangoni flow force
$\vec{F}_{\rm M, fl} = F_{\rm M, fl} \vec{e}_z$, which is by definition
the force transmitted by fluid stresses of the Marangoni flow onto
the sphere,
\begin{equation}
  F_{\rm M, fl} \equiv - \int_{S} da_i \sigma_{\rm M, iz}.
  \label{eq:FMfl}
\end{equation}
  For low Reynolds numbers, we can apply the reciprocal theorem
to the flow fields
$\vec{v}$ and $\vec{v}_{\rm M}$ and their associated stress tensors
to calculate the Marangoni flow force without explicitly
calculating the Marangoni flow $\vec{v}_{\rm M}$, as has been 
  shown in detail in Ref.\ \cite{Masoud2014}.
This  gives the identity
$0=  \int_{S+S_{\rm Int}} da_i v_{j} \sigma_{\rm M, ij}$, 
which  
leads to a Marangoni flow force
\begin{align}
  F_{\rm M,fl}
  &= -\kappa  \int_{S_{\rm Int}} dS
   \frac{\vec{v}(\vec{r})+U\vec{e}_z}{U} \cdot \vec{\nabla}_S  c(\vec{r}).
   \label{eq:FMfl2}
\end{align}
The total
Marangoni force
\begin{equation}
  F_{\rm M,tot} \equiv F_{\rm M} +  F_{\rm M,fl},
  \label{eq:FMtot}
\end{equation}
is obtained by using (\ref{FM}) and the Gauss theorem,
\begin{align}
  F_{\rm M,tot} &=  \kappa  \int_{S_{\rm Int}} dS
     \left(\vec{\nabla}_S\cdot \frac{\vec{v}(\vec{r})}{U}\right)
     c(\vec{r})\nonumber\\
 &=-\frac{3\kappa a}{2}
      \int_1^\infty d\rho \int_0^\pi d\theta \cos\theta 
    \left(\frac{1}{\rho} - \frac{1}{\rho^{3}} \right)
    {c}(\rho a,\theta)|_{y=0}.
    \label{FMtotal}
\end{align}
Because $\rho^{-1} - \rho^{-3} >0$ for $\rho>1$, 
the total Marangoni force is always positive for concentration
profiles, which are increasing  toward the rear side.
  This implies that the total Marangoni force is always
  propulsive, i.e., points in the same direction
  as the imposed velocity $U$ regardless of 
  its absolute value. This is a necessary condition for self-propulsion.
  When the particle is pulled by an external force, this also implies
  that the total Marangoni force will always support the pulling force
  instead of increasing the  drag.
The Marangoni flow contribution $F_{\rm M,fl}$, however, can have
both signs. For $F_{\rm M,fl}>0$, the flow force increases the
direct Marangoni force resulting in $F_{\rm M,tot}> F_{\rm M}$;
for  $F_{\rm M,fl}<0$, the flow force is directed backward and
increases the
drag force  resulting in $F_{\rm M,tot}< F_{\rm M}$.
As opposed to Ref.\ \cite{Lauga2012}, we will find that
both cases are possible.

In the stationary swimming state, drag and total Marangoni force
have to be balanced,
\begin{equation}
0= F_{\rm D} + F_{\rm M, tot} =  F_{\rm D} +  F_{\rm
  M} +  F_{\rm M,fl},
\label{eq:swimcond0}
\end{equation}
such that the swimmer is force-free.
This is the swimming condition that finally
determines the actual swimmer velocity $U = U_{\rm swim}$.

\subsection{Non-dimensionalization}

To proceed, we make the system of coupled equations governing our
sub-problems (i)-(iii) and the Marangoni forces dimensionless by
measuring lengths in units of $a$, velocities in units of $D/a$,
concentrations in units of $\alpha a/D$, and forces in units of $D\mu$.
We introduce
\begin{align}
  \vec{\rho} &\equiv \vec{r}/a,
               ~~\bar{\vec{\nabla}} \equiv a \vec{\nabla}
               = \vec{\nabla}_\rho,    
                     ~~\bar{\vec{v}}\equiv \vec{v} \frac{a}{D},~~
         \bar{U}  \equiv U \frac{a}{D},
 \nonumber \\
  \bar{c} &\equiv   c \frac{D}{\alpha a},~~ \bar{j} \equiv  j\frac{1}{\alpha},
 \nonumber \\
  \bar{F} &\equiv F\frac{1}{D\mu},~~ \bar{p} \equiv p \frac{a^2}{D\mu}.
            \label{eq:nondim}
\end{align}
The prescribed dimensionless velocity $\bar{U}$ of the swimmer is the first
control parameter of the problem,\footnote{In many publications on the
  diffusion-advection problem, such as Refs.\
  \cite{Acrivos1960,Acrivos1962,Acrivos1965} but also in Refs.\
  \cite{Lauga2012,Yariv2015,Vandadi2017,Boniface2019,JafariKang2020},
  $\bar{U}$ is called the Peclet number.
  Here, we define the Peclet number as ${\rm Pe} \equiv \bar{U}_{\alpha}$,
  i.e.,  by the 
  characteristic velocity $\bar{U}_{\alpha} = {\kappa \alpha a}/{D \mu}$,
  where a typical direct Marangoni force $F_{\rm M} \sim \kappa a^2
  \partial_rc(r=a)\sim \kappa a^2\alpha/D$ is balanced by the 
  typical Stokes drag force $F_D \sim \mu a U$.
  The Peclet number is a dimensionless measure of propulsion strength
  with this definition.
}
which is related to  the Reynolds number, ${\rm Re} = 2\bar{U}/{\rm Sc}$,
via the Schmidt number ${\rm Sc} \equiv\mu /\rho D$.

Our dimensionless set of equations for problems (i)-(iii) becomes 
\begin{subequations}
  \label{eq:i-iii}
\begin{align}
  {\rm (i)}&&& 
      -\left. \bar{\vec{\nabla}}  \bar{c}(\vec{\rho})\cdot \vec{n}^{\rm out}
                                     \right|_{\bar{y}=0}\approx 0\nonumber\\
    &&&\mbox{without evaporation,}   \label{eq:bcnoevap} \\
     &&&
  -\left.  \bar{\vec{\nabla}}  \bar{c}(\vec{\rho})\cdot \vec{n}^{\rm out}
                                     \right|_{\bar{y}=0}
               \approx \bar{k}\left. \bar{c}(\vec{\rho})
                            \right|_{\bar{y}=0}
                            \nonumber\\
     &&&\mbox{with evaporation,} \label{eq:bcevap} \\
{\rm (ii)}&& \bar{\vec{v}}_{\rm tot}(\vec{\rho})
                        &= \bar{\vec{v}}(\vec{\rho}) +
                                      \bar{\vec{v}}_{\rm M}(\vec{\rho}),
  \nonumber \\
{\rm (iia)}&&
  \bar{\vec{v}}(\rho,\theta) &= \bar{U}\cos\theta  u(\rho) \vec{e}_r +
                  \bar{U} \sin\theta  v(\rho) \vec{e}_\theta \nonumber\\
    &&&\mbox{Stokes~flow~field,}    \label{eq:Stokes}
  \\
{\rm (iib)}&&
  \bar{\vec{\nabla}}\cdot \bar{\vec{v}}_{\rm M} &= 0\nonumber\\
    &&&\mbox{Marangoni~flow~field,}
                             \nonumber \\
  &&\bar{\vec{\nabla}}^2 \bar{\vec{v}}_{\rm M} &=
               \bar{\vec{\nabla}} \bar{p}_{\rm M},
  \nonumber \\
  &&  \bar{\vec{v}}_{\rm M}(\infty) &=0,
                             \nonumber  \\
  &&  \left. \bar{\vec{v}}_{\rm M}(\vec{\rho}) \right|_{\rho=1}&=0,
  \nonumber \\
   &&\left. \bar{v}_{\rm M, y}(\vec{\rho}) \right|_{\bar{y}=0} &=0,
 \nonumber \\
     && \left. \partial_{\bar{y}}  \bar{\vec{v}}_{\rm M}(\vec{\rho})
        \right|_{\bar{y}=0}
              &= -{\rm Pe}
    \left.    \bar{\vec{\nabla}}_{S}  \bar{c}(\vec{\rho}) \right|_{\bar{y}=0},
                          \label{eq:Mflow}\\      
{\rm (iii)}&&  0 &=  \bar{\vec{\nabla}}^2 \bar{c} -
               (\bar{\vec{v}}(\vec{\rho})+ \bar{\vec{v}}_{\rm M}(\vec{\rho}))
      \cdot \bar{\vec{\nabla}} \bar{c},
  \nonumber \\
 &&   \bar{c}(\infty)  &= 0,
   \nonumber  \\
 &&
        \left. \bar{\vec{j}}\cdot\vec{n}\right|_S
             &=-
               \left. \bar{\vec{\nabla}} \bar{c}\cdot\vec{n} \right|_S
               =1,
    \label{eq:boundconstflux}
\end{align}
\end{subequations}
with the dimensionless Peclet number
\begin{align}
 {\rm Pe} &\equiv \frac{\kappa \alpha a^2}{D^2 \mu}=
             \frac{\kappa \dot{m}}{2\pi D^2 \mu}, 
            \label{eq:Pe}
 \end{align}
where $\dot{m}= 2\pi a^2\alpha$ is the mass loss per time
of the swimmer (see Fig.\ \ref{fig:mdot}).
The Peclet number is a dimensionless measure of propulsion strength.

Typical values for the PEG-alginate swimmer are very high, ${\rm Pe} \sim
10^7$ (see Table \ref{tab:nondimen}). 
We also introduced the  dimensionless Biot number
\begin{equation}
   \bar{k} \equiv \frac{ak}{D}
   \label{eq:Biot}
\end{equation}
governing possible evaporation, which is practically absent for PEG.
From Eq.\ (\ref{eq:Mflow}), we see that the Peclet number ${\rm Pe}$
determines the velocity scale of the Marangoni flow field.
Therefore, we can also assign a Reynolds number
${\rm Re}_{\rm M} = {2{\rm Pe}}/{{\rm Sc}}= {\rm Re} {\rm Pe}/\bar{U}$
to the Marangoni flow. 
Typical values for the PEG-alginate swimmer are  ${\rm Re}_{\rm M}  \sim
10^4$ (see Table \ref{tab:nondimen}), which agrees with the experimentally 
observed  turbulent features of Marangoni flows (see Fig.\ \ref{fig:PIV}). 
Via the advection with $\bar{\vec{v}}(\vec{\rho})+ \bar{\vec{v}}_{\rm
  M}(\vec{\rho})$,  the concentration field $c(\vec{\rho})$ depends
both on the dimensionless velocity scale $\bar{U}$ of the Stokes field
and the dimensionless velocity scale ${\rm Pe}$ of the Marangoni
flow field, in general.
All dimensionless parameters are summarized in Table \ref{tab:nondimen},
along with typical values for the PEG-alginate swimmers and in comparison with
camphor boats according to Ref.\ \cite{Boniface2019}.

\begin{table*}
  \begin{center}
  \caption{\label{tab:nondimen}
    Dimensionless parameters.
    ${\rm Re}$ or $\bar{U}$,
    ${\rm Sc}$, ${\rm Pe}$ and $\bar{k}$ are control parameters
    of the problem.  ${\rm Re}_{\rm M}$ and ${\rm Nu}$
    cannot be independently  controlled but characterize the
    resulting solutions; the swimming velocity $\bar{U}_{\rm swim}$
    is determined by the
     force balance swimming condition.} 
   \begin{tabular}{ l l @{\qquad} | l | l | l}
     \hline\noalign{\smallskip}
     Parameter &      Formula      & Eq     & 
                                             PEG-alginate swimmer
     & Camphor   boat\cite{Boniface2019}
     \\
   \noalign{\smallskip}\hline\noalign{\smallskip}
   Reynolds number ${\rm Re}$ &  $ ={2\rho U a}/{\mu}
                  = {2\bar{U}}/{{\rm Sc}}$  &    &
                                      $30-80$ &  $60-3000$ \\
   dimensionless velocity  $\bar{U}$  & $ = U{a}/{D}  $
                        & (\ref{eq:nondim}) & $4\times 10^4-1.2\times 10^5$
                                      &  $4\times 10^4-1.2\times 10^6$ \\
   Schmidt number ${\rm Sc}$  &  $ = {\mu}/{\rho D}$  & 
                                         & $2860$ &  $1390$ \\
   Peclet number ${\rm Pe}$ & $ =  {\kappa \alpha a^2}/{D^2 \mu}$
                                  &    (\ref{eq:Pe}) &
                    $3.5\times 10^{6} - 8.8\times 10^{7}$ & 
             $(9.3\times 10^{9})(a/4{\rm mm})^2$\\
     Biot number $\bar{k}$ & $= {ak}/{D}$ & (\ref{eq:Biot}) &
                               $\approx 0$ & $\approx   550$
     \\
     \noalign{\smallskip}\hline\noalign{\smallskip}
   swimming velocity  $\bar{U}_{\rm swim}$  & $ = U_{\rm swim}{a}/{D}  $
                        & (\ref{eq:swimcond}) &  $4\times 10^4-1.2\times 10^5$
                                      &  $4\times 10^4-1.2\times 10^6$ \\
   Marangoni Reynolds number ${\rm Re}_{\rm M}$ &
   $= {2{\rm Pe}}/{{\rm Sc}}= {\rm Re} {\rm Pe}/\bar{U}$ &  & $2.4\times 10^3-6.2\times 10^4$ &
                         $(1.4\times 10^7) (a/4{\rm mm})^2$
     \\
   Nusselt (Sherwood) number ${\rm Nu}$ (${\rm Sh}$)
            &   $ ={-\partial_\rho \bar{c}_0(1)}/{\bar{c}_0(1)}$ &
                    (\ref{eq:Nu}) &  \multicolumn{2}{l}{(\ref{eq:NuU})} \\
\noalign{\smallskip}\hline
   \end{tabular}
   \end{center}
\end{table*}

In the following, we will  solve the problems (i)-(iii), in order to  obtain
the  dimensionless direct and total Marangoni forces [see Eqs.\ (\ref{FM}] and
(\ref{FMtotal}))
from the concentration profiles by
\begin{align}
  \bar{F}_{\rm M} 
  &= - 2 {\rm Pe} \int_0^\pi d\theta \cos\theta  \bar{c}(1,\theta)|_{\bar{y}=0},
                \label{FMdim}\\
 \bar{F}_{\rm M,tot} 
  &=-\frac{3{\rm Pe}}{2}
      \int_1^\infty d\rho \int_0^\pi d\theta \cos\theta 
    \left(\frac{1}{\rho} - \frac{1}{\rho^{3}} \right)
    \bar{c}(\rho,\theta)|_{\bar{y}=0}.
    \label{FMtotaldim}
\end{align}
for a prescribed swimmer velocity $\bar{U}$. 
Finally, force balance gives the dimensionless
version of the swimming condition (\ref{eq:swimcond0}), 
\begin{equation}
  -\bar{F}_{\rm D} = 3\pi  \bar{U}_{\rm swim} =
  \bar{F}_{\rm M}({\rm Pe},\bar{U}_{\rm swim}) +
  \bar{F}_{\rm M,fl} ({\rm Pe},\bar{U}_{\rm swim}),
  \label{eq:swimcond}
\end{equation}
which then selects the actual swimmer velocity $\bar{U}=\bar{U}_{\rm swim}$
as a function of the remaining control parameters
${\rm Pe}$ (``fuel'' emission) and eventually $\bar{k}$ (evaporation). 

The non-dimensionalization reveals that the coupled problems (i)-(iii)
and the Marangoni forces 
depend on three dimensionless control parameters (see also table
\ref{tab:nondimen}):
First, the prescribed dimensionless velocity of the swimmer $\bar{U}$;
second, the Peclet number $\rm Pe$ characterizing the strength $\alpha$
of the  surfactant emission, and third, the Biot number $\bar{k}$
characterizing the evaporation. 
We also see that the Peclet number both controls the
strength of the Marangoni flow via Eq.\ (\ref{eq:Mflow}) and
the strength of all  Marangoni forces. 
We note, however,
that $\bar{F}_{\rm M}/{\rm Pe}$ and $\bar{F}_{\rm M,tot}/{\rm Pe}$
still depend on $\bar{U}$ and
${\rm Pe}$ via the dependence of  $\bar{c}(\vec{\rho})$ on these
parameters.

Another important finding from non-dimensionalization is that 
the diffusion-advection problem decouples from the Marangoni flow
problem for ${\rm Pe} \ll \bar{U}$, where we can
neglect   $\vec{v}_{\rm M}$ in the advection term. 
Then, the concentration profile is only determined
by Stokes flow,  becomes axisymmetric, and  only depends on $\bar{U}$.
In this limit,  
the Marangoni flow field need not to be calculated in order
to calculate the total Marangoni force via Eq.\ (\ref{FMtotaldim}).

\subsection{Numerical methods}

Numerically, we only address the low Reynolds number regime.
In general, we consider the problems (i)-(iii), i.e.,
solve the coupled diffusion-advection problem and the Marangoni flow problem
for a prescribed swimmer velocity $\bar{U}$. From the
solution for the concentration field, we then calculate the
Marangoni forces as a function of $\bar{U}$ and ${\rm Pe}$
in order to finally solve the force balance
swimming condition. 
We use an iterative  finite element
scheme to solve the full coupled problem;
this approach is explained in detail in Appendix \ref{app:numerical}.

\subsection{Low Reynolds number results}

Low Reynolds numbers ${\rm Re}\ll 1$
are realized for $\bar{U} \ll {\rm Sc}/2$, which can still be much larger
than unity as typical Schmidt numbers for surfactants in aqueous
solutions are of the order of $1000$ (see Table \ref{tab:nondimen}).
Therefore, we have to discuss \emph{both} the diffusive limit $\bar{U}\ll 1$
\emph{and} the advective limit $\bar{U}\gg 1$.


\subsubsection{Decoupled limit ${\rm Pe} \ll \bar{U}$}

First, we will consider the  limit 
${\rm Pe} \ll \bar{U}$, where the diffusion-advection problem
for a half-sphere with prescribed velocity $U$
decouples from the Marangoni flow problem.
We also focus on the case 
in the absence of evaporation first, as it is appropriate for the
PEG-alginate swimmer.

Diffusive release from a moving emitter or from a resting emitter in a
fluid flow can be  characterized by the
average Nusselt number (or Sherwood number {\rm Sh}),
\begin{equation}
  {\rm Nu} \equiv \frac{\int_S \vec{j}(\vec{r})\cdot\vec{n}\,dA}
  { (D/a) \int_S c(\vec{r})\,dA},
  \label{eq:Nu}
\end{equation}
which is the dimensionless ratio of the total emitted flux  and the
typical diffusive flux \cite{Leal}. A Nusselt number of one is realized for
purely diffusive transport, Nusselt numbers much larger than one indicate
strong advective transport.

\begin{figure}
  \begin{center}
    \includegraphics[width=0.99\linewidth]{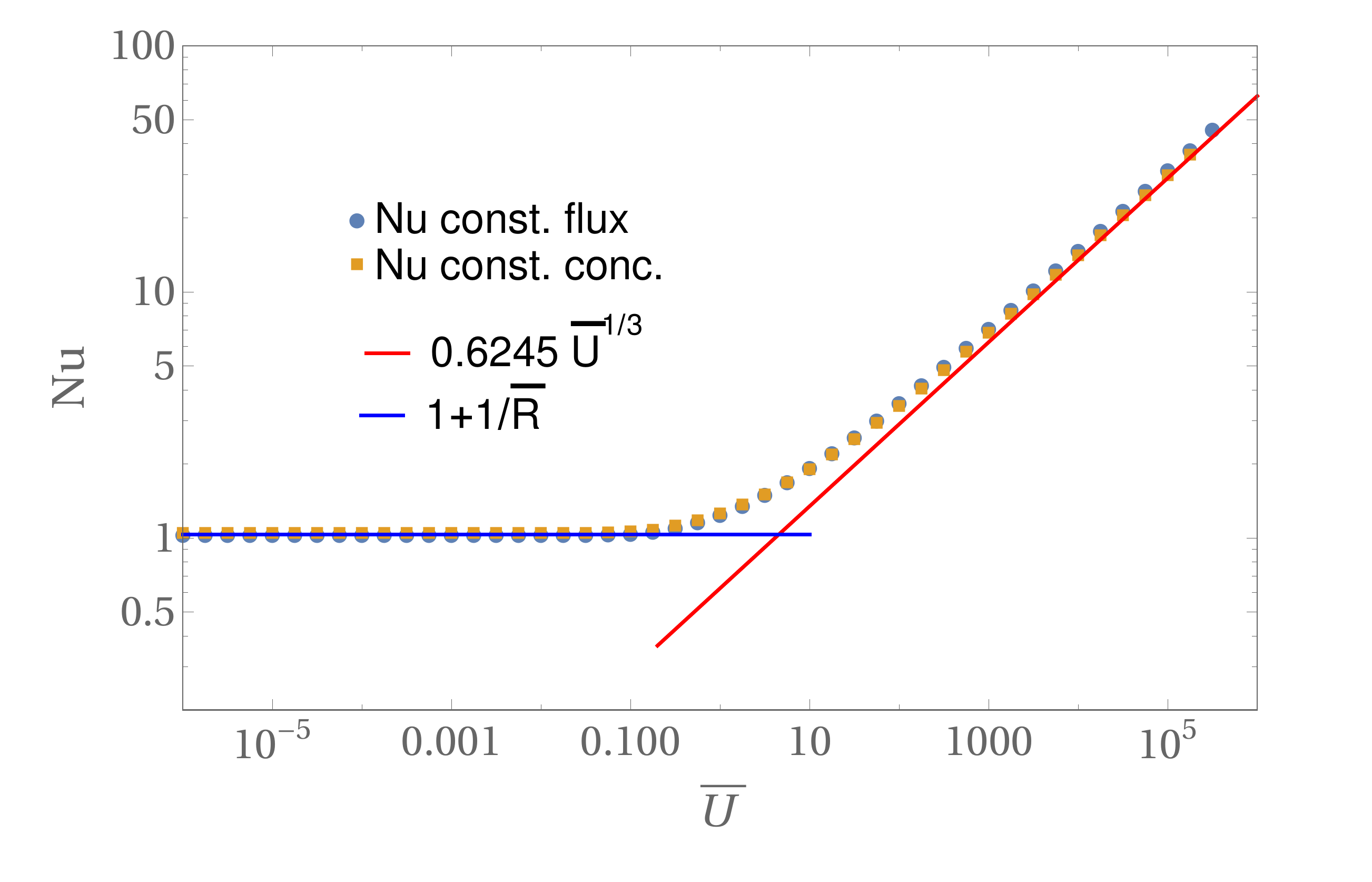}
   \caption{\label{fig:Nusselt}
     Average Nusselt number as a function of $\bar{U}$
     for constant flux and constant
     concentration boundary conditions
     in the decoupled limit  ${\rm Pe} \ll \bar{U}$.
       All results are from numerical FEM solutions of the axisymmetric 
   diffusion-advection
   equation in two-dimensional
   angular representation with $\rho<\bar{R}=30$.
 }
 \end{center}
 \end{figure}

In the decoupled limit 
  ${\rm Pe} \ll \bar{U}$, we find for the Nusselt number 
\begin{align}
  {\rm Nu} = \frac{1}{\bar{c}_0(\rho=1)}&=
\begin{cases}
   1+\frac{1}{2} \tilde{U} & \mbox{for}~\bar{U}\ll 1\\
   0.65\, \bar{U}^{1/3}
   & \mbox{for}~\bar{U}\gg 1
 \end{cases},
     \label{eq:NuU}
\end{align}
where $c_0(\rho) \equiv \frac{1}{2} \int_0^\pi d\theta \sin\theta 
\bar{c}(\rho,\theta)$ is the zeroth Legendre coefficient.
There are  two regimes, a diffusive regime
  for $\bar{U}\ll 1$ characterized by a Nusselt number close to
  one and an advective regime for $\bar{U} \gg 1$ where
  the Nusselt number becomes large, which is also
  clearly supported by the numerical results in Fig.\
  \ref{fig:Nusselt}. 
  The result (\ref{eq:NuU}) can be derived analytically \cite{Ender2020},
  apart from the value of the prefactor  
 $0.65$, which we determined numerically from the
 data in Fig.\ \ref{fig:Nusselt}. A short derivation based on scaling
 arguments for the advective regime is presented below. 
 The numerical results in Fig.\
\ref{fig:Nusselt} show perfect agreement and clearly confirm
the existence of just two regimes.

The Nusselt number has been originally defined for constant
concentration boundary conditions $\bar{c}(1,\theta)=1$,
For constant concentration boundary conditions,
the result
is well-known \cite{Acrivos1960,Acrivos1962,Acrivos1965,Leal}
and very similar (see Fig.\ \ref{fig:Nusselt}),
\begin{align}
{\rm Nu} = -\partial_\rho\bar{c}_0(\rho=1)&=
\begin{cases}
   1+\frac{1}{2} \tilde{U} + ... & \mbox{for}~\bar{U}\ll 1\\
   0.6245\, \bar{U}^{1/3}
   & \mbox{for}~\bar{U}\gg 1
 \end{cases}
 \label{eq:NuUB}
\end{align}
with a prefactor that can be calculated analytically. 
This indicates that literature results for the Nusselt number
for constant concentration boundary conditions also apply
to our situation of constant flux boundary conditions,
which is an important insight that we will assume
to also hold for higher Reynolds numbers below.

\begin{figure}
  \begin{center}
    \includegraphics[width=0.99\linewidth]{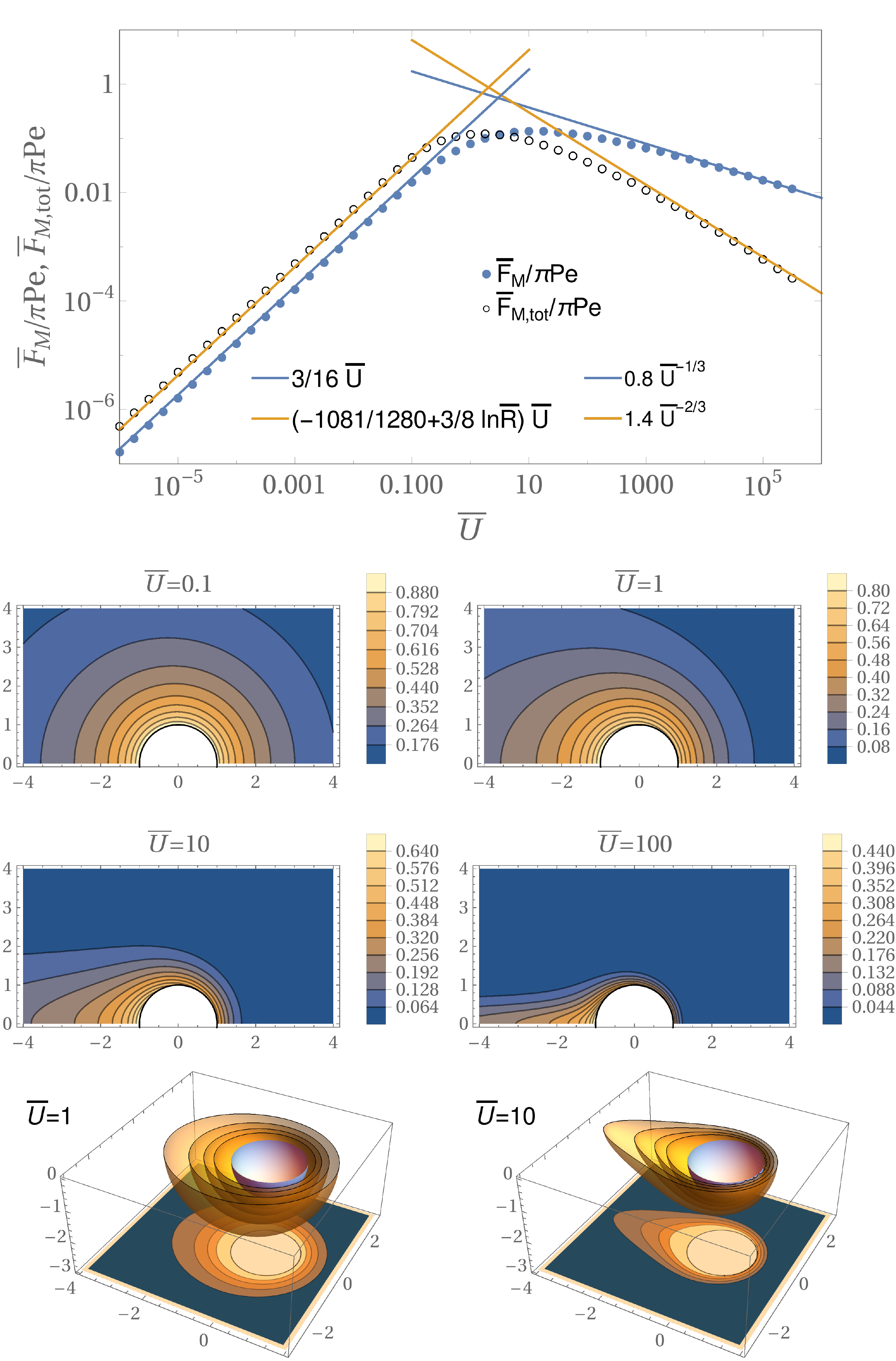}
   \caption{\label{fig:FM}
     Marangoni forces ${\bar{F}_{\rm M}}/{\pi \rm Pe}$ and
     ${\bar{F}_{\rm M, tot}}/{\pi \rm Pe}$ as a function of imposed
       velocity $\bar{U}$
     together with corresponding 
     concentration profiles 
     (in the $\bar{z}\bar{x}$-plane and in three dimensions)
     in the decoupled limit  ${\rm Pe} \ll \bar{U}$. 
   All results are from numerical FEM solutions of the axisymmetric 
   diffusion-advection
   equation in two-dimensional
   angular representation with $\rho<\bar{R}=30$.
   In the advective regime $\bar{U}\gg 1$ a concentration boundary
     layer develops [see Eq.\ (\ref{eq:Deltarho})].
 }
 \end{center}
\end{figure}

 In the decoupled limit  ${\rm Pe} \ll \bar{U}$, we also find
  a diffusive and an advective regime for the 
 Marangoni forces 
\begin{align}
 \frac{\bar{F}_{\rm M} }{\pi {\rm Pe}}  &=
\begin{cases}
   \frac{3}{16} \bar{U} & \mbox{for}~\bar{U}\ll 1\\
   d_{\rm M} \bar{U}^{-1/3}~\mbox{with}~d_{\rm M}\simeq 0.8
   & \mbox{for}~\bar{U}\gg 1
 \end{cases},
     \label{eq:c1Unew}
  \\
  \frac{\bar{F}_{\rm M, tot}}{\pi \rm Pe}
                    &=
       \begin{cases}
   -  \frac{1081}{1280}  \bar{U}
          +\frac{3}{8}  \bar{U}\ln \bar{R} & \mbox{for}~\bar{U}\ll 1\\
   d_{\rm M, tot} \bar{U}^{-2/3}~\mbox{with}~d_{\rm M, tot}\simeq 1.4
   & \mbox{for}~\bar{U}\gg 1
 \end{cases},
     \label{eq:c1Mflnew}
\end{align}
where numerical constants $d_{\rm M}$ and $d_{\rm M, tot}$ are
obtained from the numerical results, see Fig.\ \ref{fig:FM},
and $\bar{R}$ is the radial system size.
  Again,  the numerical results (Fig.\
\ref{fig:FM}) show perfect agreement and clearly confirm
the existence of just two regimes, a diffusive and an advective regime.
Direct and total Marangoni force reach maximal values 
$\bar{F}_{\rm M},\bar{F}_{\rm M, tot}\sim 0.15\pi{\rm Pe}$
in the crossover region 
 $\bar{U} \sim 1$ between diffusive and advective transport.

   Also, the results (\ref{eq:c1Unew}) and (\ref{eq:c1Mflnew})
   can be derived analytically from a calculation of
   the concentration field \cite{Ender2020},
   apart from the value of the numerical constants.
 Here we present a short derivation based on scaling
 arguments.
In the diffusive limit  $\bar{U}\ll 1$, there is a linear response of the
concentration field, which leads to a linear response
of the Nusselt number and Marangoni forces.
The coefficients can be calculated by perturbation theory
about the concentration field $c^{(0)}(\vec{\rho}) = 1/\rho$ at
$\bar{U}=0$ in powers of $\bar{U}$.
A remarkable result of this calculation is that the linear term
for the total Marangoni force diverges logarithmically with the system
size $R$, see Eq.\ (\ref{eq:c1Mflnew}), while the linear term for the
direct Marangoni force stays finite. This means that the Marangoni
flow forces strongly \emph{increase} the direct force for $\bar{U}\ll 1$;
such a behavior could not be found in
Ref.\ \cite{Lauga2012}.
For very large system sizes $\bar{R}\gg 1/\bar{U}$,
the large scale cutoff $\bar{R}$ in (\ref{eq:c1Mflnew})
will be replaced  by $1/\bar{U}$ because the convection term
can no longer be treated perturbatively in the region $\rho\gg 1/\bar{U}$,
regardless how small $\bar{U}$ is \cite{Acrivos1962}.
  We also
  note that the result ${\rm Nu} \approx  1+\frac{1}{2} \tilde{U}$
   from Eqs.\ (\ref{eq:NuU}) and (\ref{eq:NuUB})
    for the Nusselt number in the diffusive regime $\bar{U}\ll 1$ 
    is derived from the non-perturbative matching procedure
    for very large system sizes $\bar{R}\gg 1/\bar{U}$ \cite{Acrivos1962},
  while a perturbative  calculation gives
  ${\rm Nu} \approx  1+1/\bar/{R}+ O(\bar{U}^2)$
  with the radial system size
  $\bar{R}$. This  perturbative result
  describes our numerical data for  a finite system
  actually better, see Fig.\ \ref{fig:Nusselt}.

In the limit of strong advection
$\bar{U}\gg 1$,
a concentration boundary layer develops around the
  half-sphere, as can be clearly
seen in the  concentration
profiles  in Fig.\ \ref{fig:FM}).
Its width $\Delta r$ is determined by the distance that
a surfactant molecule can diffuse during the time
$\Delta t\sim a/v(\Delta r/a)$ [see Eq.\ (\ref{v})]
it takes to be transported along the
sphere by advection: $\Delta r^2 \sim D \Delta t$. Because
$v(\Delta r/a) \sim  U \Delta r/a$  [see Eq.\ (\ref{v})], we find
\begin{equation}
  \Delta \rho = \Delta r/a  \sim  \bar{U}^{-1/3}.
  \label{eq:Deltarho}
\end{equation}
This is a classic result for the diffusion-advection problem
for constant concentration boundary conditions \cite{Acrivos1960,Leal},
but also holds for constant flux boundary conditions. 
Because  the concentration will drop in radial direction
  from its value at the surface $S$ of the half-sphere  to zero
  within the concentration
boundary layer of width $\Delta \rho$, we also have
$1= -\partial_\rho \bar{c}(\rho=1,\theta)\sim \bar{c}(\rho=1)/\Delta \rho $,
which leads to a scaling
\begin{equation}
  \bar{c}(\rho=1,\theta) \sim \Delta \rho \sim \bar{U}^{-1/3}
  \label{eq:cDeltarho}
\end{equation}
of the symmetry-breaking concentration level at the surface $S$ of the
  half-sphere
for constant flux boundary conditions.
For strong advection, the Marangoni forces decrease as a function
of $\bar{U}$ because the concentration  boundary layer width $\Delta
  \rho$,
in which forces are generated, 
and the concentration
level around the sphere, to which forces are
proportional, both decay with velocity as $\bar{U}^{-1/3}$.

The scaling property (\ref{eq:cDeltarho}) for $\bar{c}(\rho=1,\theta)$
directly explains the results (\ref{eq:NuU}),
${\rm Nu} \sim 1/\bar{c}(\rho=1,\theta) \sim \bar{U}^{1/3}$,
for the Nusselt number and (\ref{eq:c1Unew}),
$ \bar{F}_{\rm M}/{\rm Pe}\sim  \bar{c}(\rho=1,\theta) \sim \bar{U}^{-1/3}$,
for the
direct Marangoni force in the advective limit $\bar{U}\gg 1$.
They are clearly confirmed by all numerical results in Figs.\
  \ref{fig:Nusselt} and \ref{fig:FM}.
The result for the total Marangoni force (\ref{eq:c1Mflnew})
seems to deviate from this advective scaling
Here, the expected scaling from the radial concentration
  boundary layer of width $\Delta \rho$ 
  is $ \bar{F}_{\rm M,tot}/{\rm Pe}\sim \Delta \rho^2 
 \bar{c}(\rho=1,\theta) \sim \bar{U}^{-1}$ (see Eq.\ (\ref{FMtotaldim})),
which is clearly not found in the numerics (yellow line in 
Fig.\ \ref{fig:FM}).
The reason is that 
 this contribution is actually
 only sub-dominant.  The leading contribution comes from
 the advective tail  of angular width $\Delta \theta \sim
 \bar{U}^{-1/3}$;
 the width of the tail follows from the scaling of the 
 stream function $\psi \propto r^2\Delta \theta^2$ in the tail and $\psi 
\propto  3\Delta r^2\sin^2\theta/2$ in the boundary layer
and the fact that a fluid particle should follow a Stokes flow 
stream line 
$\psi= {\rm const}$ in the advective limit.
Therefore, the dominant contributions in Eq.\ (\ref{FMtotaldim})
are  $ \bar{F}_{\rm M,tot}\sim {\rm Pe}  \Delta \theta
\bar{c}(\rho=1,\theta)\sim \bar{U}^{-2/3}$
in agreement with the numerical results in Fig.\ \ref{fig:FM}.

Comparing the curves for direct and total Marangoni force in 
  Fig.\ \ref{fig:FM}, we observe a crossing such that 
${\bar{F}_{\rm M}} < {\bar{F}_{\rm M, tot}}$ 
in the diffusive regime $\bar{U}\ll 1$, while
${\bar{F}_{\rm M}} > {\bar{F}_{\rm M, tot}}$ in the advective regime
$\bar{U} \gg 1$. This means that the Marangoni flow force 
$\bar{F}_{\rm M,fl} =  \bar{F}_{\rm M,tot} - \bar{F}_{\rm M}$
\emph{increases} the propulsion
force in the diffusive regime  $\bar{U} \ll 1$ but \emph{decreases}
the propulsion force (or increases the drag) for $\bar{U} \gg 1$.
This subtle result is related to the structure of the Marangoni flows,
which are generated by the surfactant concentration gradients 
$ \left.\bar{\vec{\nabla}}_{S}  \bar{c}(\vec{\rho}) \right|_{\bar{y}=0}$
within the liquid-air interface (see Eq.\ (\ref{eq:Mflow}))
 and can  be qualitatively rationalized with the help of Eq.\
 (\ref{eq:FMfl2}) for the Marangoni flow force.

\begin{figure}
  \centerline{\includegraphics[width=0.99\linewidth]{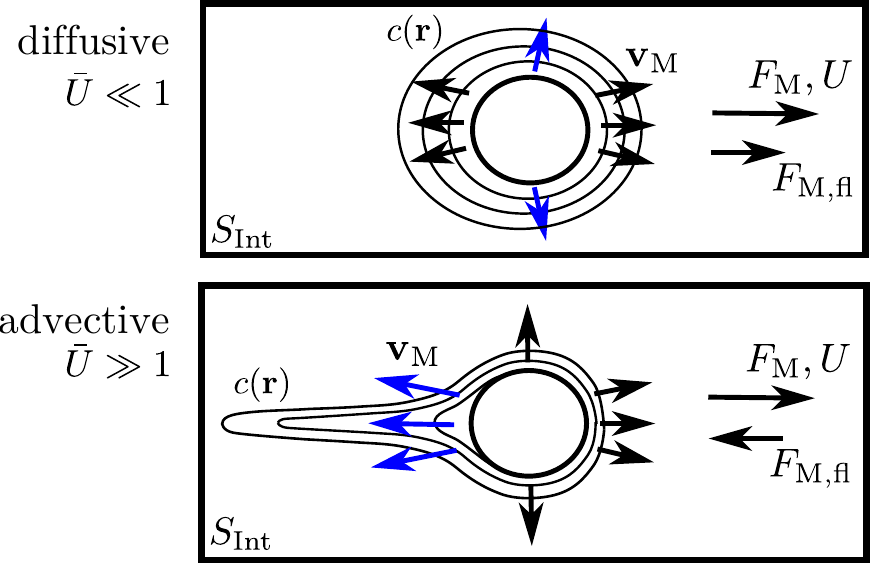}}
  \caption{Schematic  of concentration field (as concentration
    contour lines) and Marangoni flow 
    field (arrows)
    and resulting direct force $F_{\rm M}$ and Marangoni flow force 
$F_{\rm M,fl}$ in the diffusive ($\bar{U}\ll 1$) and advective ($\bar{U}\gg
1$) regime. The dominant Marangoni flow contributions are marked in blue.
In the diffusive regime, Marangoni flows increase the direct force,
in the advective regime they decrease the direct force.
It is important to note that concentration contour lines are
  not approaching a spherical shape close to the swimmer in the advective
  regime because we
  have constant flux boundary conditions such that the
  swimmer surface is a contour of constant radial gradient
  by construction.
  The elongated tail gives rise to a slower
decay of the gradient itself and, thus, larger radial gradients
in the tail, which is reflected by additional contour lines
emerging on the rear side.
     }
\label{fig:scheme_flow}
\end{figure}

  We can decompose
  the surfactant concentration gradient into tangential and radial components
  \begin{equation}
    \left.\bar{\vec{\nabla}}_{S}  \bar{c}(\vec{\rho})
    \right|_{\bar{y}=0} = \frac{1}{\rho}\partial_\theta \bar{c}\, \vec{e}_\theta +
    \partial_\rho\bar{c} \,\vec{e}_r.
  \end{equation}
Because of   advection 
the  tangential  component 
 points from the front  to the rear corresponding to an increasing 
 surfactant concentration toward the rear ($\partial_\theta \bar{c}>0$).
 It gives rise to a forward Marangoni flow and $F_{\rm M,fl}>0$ in Eq.\
 (\ref{eq:FMfl2})
 because $ -\vec{e}_z \cdot \rho^{-1}\partial_\theta \bar{c}
 \vec{e}_\theta \propto \sin\theta >0$.
This is the dominating effect in the diffusive regime $\bar{U} \ll 1$,
where the perturbation theory gives to 
leading linear order in $\bar{U}$ a contribution of the form
  $\bar{U}\bar{c}^{(1)} \propto -\bar{U}\bar{c}_1(\rho)
  \cos\theta$ resulting in
  $\rho^{-1}\partial_\theta \bar{c}=
  \bar{U}\rho^{-1}\bar{c}_1(\rho) \sin\theta>0$.
  The front-directed tangential Marangoni flow components give
  rise to the blue flow directions in the schematic in
  Fig.\ \ref{fig:scheme_flow} (top).

The radial component 
points inward ($ \partial_\rho\bar{c} <0$) because of the radially decaying
surfactant concentration. This  gives rise 
to radially outward Marangoni flows. 
Because  $ -\vec{e}_z \cdot \partial_\rho\bar{c} \vec{e}_r\propto
\cos\theta$ in Eq.\
 (\ref{eq:FMfl2}),  this increases
the direct force in the
front  (around $\theta=0$) but decreases it in the back (around $\theta=\pi$).
Advection
 leads to bigger surfactant concentrations in the rear, which  also result in
bigger radial
concentration gradients on the rear side and lead
to an overall  decrease in the direct force $\bar{F}_{\rm M,fl}<0$
and, thus,  an increased drag.
This is the dominating effect in the advective regime $\bar{U}\gg 1$,
where a concentration boundary layer of width $\Delta \rho \sim \bar{U}^{-1/3}$
forms around the half-sphere,
which results in steep radial concentration gradients that
are stronger on the rear side.
The stronger radial  Marangoni flow components in the rear 
are indicated by the blue arrows in Fig.\ \ref{fig:scheme_flow} (bottom);
this phenomenon is also visible in the experimental PIV measurements
in Fig.\ \ref{fig:PIV}(B) during motion.
There are also slightly bigger
radial concentration gradients in the diffusive regime
$\bar{U}\ll 1$, but they are sub-dominant for  the slow
radial decay of  the function $\bar{c}_1(\rho)$ in the absence of a
radial concentration boundary layer.

\subsubsection{Strong Marangoni flow ${\rm Pe} \gg \bar{U}$}
\label{sec:strongM}

For weak Marangoni flow, ${\rm Pe} \ll \bar{U}$, we could
  decouple the diffusion-advection problem and obtained to regimes,
  a diffusive or linear response regime for $\bar{U}\ll 1$ and an
  advective regime for $\bar{U}\gg 1$. Now we increase
  the Peclet number ${\rm Pe}$ and, thus, the Marangoni flow.
For a strong Marangoni flow, ${\rm Pe} \gg \bar{U}$, the linear
response regime $\bar{U}\ll 1$ becomes modified.
We first have to consider the dominant Marangoni flow problem (iib),
which determines the main component of the fluid flow 
in the diffusion-advection problem (iii). 
The Marangoni  flow pattern is a  stationary
 Marangoni vortex ring 
 around the spherical swimmer  below and parallel to
 the fluid interface $S_{\rm Int}$.
 Because this solution lacks axisymmetry an analytical solution is no longer
 possible. Nevertheless, we can obtain novel scaling results
 for concentration profile and Marangoni flow field
 in a concentration boundary layer of width $l_c$ below
 the fluid interface $S_{\rm Int}$
 along similar lines as Refs.\ \cite{LeRoux2016,Roche2014}.

 The surfactant is emitted from the sphere with the
 fixed current density $\left.\bar{j}\right|_S=1$.
 Advection by the Marangoni flow  $\bar{\vec{v}}_{\rm M}$
 concentrates the surfactant in the boundary layer
 of width $\bar{l}_c$ below  $S_{\rm Int}$. It takes a time 
 $t \sim r/v_{\rm M}$  to reach a radial distance $r$.
 During this time, the surfactant diffuses over a distance
 $l_c \sim (D t)^{1/2} \sim (Dr/v_{\rm M})^{1/2}$ 
 or
 $\bar{l}_c \sim  (\rho/\bar{v}_{\rm M})^{1/2}$ in vertical
   $y$-direction, which sets the
 boundary layer width $\bar{l}_c$. Because we are at low Reynolds numbers,
 the laminar boundary layer below the fluid interface $S_{\rm Int}$
 is  of the size  $\delta \sim a$
 ($\bar{\delta} \sim 1$)  set by the sphere.
 The laminar boundary layer governs the decay of the Marangoni flow field
   $v_{\rm M}$ in  $y$-direction.
 
 Moreover, we have mass conservation of the emitted surfactant.
 The total
 advective flow
 $\bar{J} \sim 2\pi \bar{c} \bar{v}_{\rm M}\rho  \bar{l}_c$ 
 below  the interface at distance
 $\rho$ will always equal the original flow $\bar{J}=2\pi$ that
 is emitted at the half-sphere,
 \begin{equation}
   1 = \bar{J}/2\pi \sim \bar{c} \bar{v}_{\rm M}\rho  \bar{l}_c
   \sim \bar{c} \bar{v}_{\rm M}^{1/2}\rho^{3/2} .
   \label{eq:jadv}
 \end{equation}
 In addition, the Marangoni boundary condition (see Eq.\ (\ref{eq:Mflow})
 gives a second equation 
 \begin{equation}
   - {\rm Pe} \partial_\rho\bar{c} =  \partial_{\bar{y}} \bar{v}_{\rm M}
   \sim \frac{\bar{v}_{\rm M}}{\bar{\delta}} \sim \bar{v}_{\rm M}
   \label{eq:cvM}
 \end{equation}
 for concentration and velocity, which follows from the definition
    of the laminar boundary layer width $\bar{\delta}$.
Combining both Eqs.\ (\ref{eq:jadv}) and (\ref{eq:cvM}), we find
a  differential equation for $\bar{c}(\rho)$, which
we solve with the  boundary condition $\bar{c}(\infty)=0$
resulting in
\begin{align}
  \bar{c}(\rho) &= \bar{c}(1) \rho^{-2/3}
  ~~\mbox{with}~~ \bar{c}(1) \sim {\rm Pe}^{-1/3},
  \label{eq:cMarangoni}\\
  \bar{v}_{\rm M} &\sim \bar{c}^{-2} \rho^{-3}
  \sim \bar{c}^{-2}(1) \rho^{-5/3} \sim {\rm Pe}^{2/3} \rho^{-5/3}.
  \label{eq:vMarangoni}
\end{align}
We see that the advective current
$\bar{j}_{\rm M} \sim \bar{c} \bar{v}_{\rm M}\sim {\rm Pe}^{1/3} \rho^{-7/3}$
becomes smaller than the corresponding diffusive current
$\bar{j}_D \sim -\partial_\rho \bar{c} \sim {\rm Pe}^{-1/3} \rho^{-5/3}$
for $\rho > {\rm Pe}$. Then our assumption of advective transport
breaks down, and this should mark the boundary of the
Marangoni advection dominated region. Therefore,
$\rho_{\rm M} \sim  {\rm Pe}$
should be the scaling of the size of the Marangoni vortex
around the sphere for low Reynolds numbers.
At larger distances, a crossover to diffusive transport
with  $\bar{c} \propto \rho^{-1}$  sets in.

So far, we considered the leading order of our problem by setting
$\bar{U}\approx 0$; going one order further, 
we get the linear response for small $\bar{U}$
with the ansatz
$\bar{c} =  \bar{c}^{(0)}   + \bar{U} \bar{c}^{(1)}$
with $\bar{c}^{(0)}(\rho)$ given by (\ref{eq:cMarangoni}).
In the total flow $\vec{v} + \vec{v}_{\rm M}$, the
Marangoni flow (\ref{eq:vMarangoni}) is the zeroth order result,
$\vec{v}_{\rm M}= \vec{v}_{\rm M}^{(0)}$, 
while the Stokes swimming flow $\vec{v}= \vec{v}^{(1)}$
is linear in $\bar{U}$.
In an advection dominated situation, the 
constant flux relation (\ref{eq:jadv}) still holds in the
presence of Stokes flow,
\begin{equation}
  1 \sim  (\bar{c}^{(0)} + \bar{U} \bar{c}^{(1)})
  (\bar{U}\bar{u}\cos\theta + \bar{v}_{\rm M})^{1/2}\rho^{3/2},
 \end{equation}
 where $\bar{u}(\rho)$ is the radial component  of the Stokes flow. 
Expanding up to first order in $\bar{U}$ and using
 (\ref{eq:jadv}) for the leading order, we find
 \begin{equation}
  \bar{c}^{(1)}(\rho) \sim   \frac{1}{ \bar{v}_{\rm M}^{1/2}(\rho)}
                        \bar{c}^{(0)}(\rho)  \bar{u}(\rho)
                        \sim {\rm Pe}^{-2/3}  \rho^{1/6} \bar{u}(\rho).
 \end{equation}
 This contribution is symmetry-breaking; inserting this scaling
 of  the concentrations into the relations
 (\ref{FMdim}) and (\ref{FMtotaldim} for the 
 Marangoni forces, we obtain  
 \begin{align}
   \frac{ \bar{F}_{\rm M}}{\pi{\rm Pe}}
   &\sim \bar{U}  {\rm Pe}^{-2/3},&
   \frac{ \bar{F}_{\rm M,tot}}{\pi {\rm Pe}}
   &\sim \bar{U}  {\rm Pe}^{-2/3}.
   \label{eq:FMtotalMarangoni}
 \end{align}
 We checked these predictions numerically in Fig.\ \ref{fig:FMtotalPe}
   by using our iterative FEM approach (see Appendix \ref{app:numerical}),
   which is possible up to ${\rm Pe}\sim
   50$ and find good agreement, in particular, for the predictions
   $\bar{F}_{\rm M}/{\rm Pe} \propto  {\rm Pe}^{-2/3}$ and
   $\bar{F}_{\rm M,tot}/{\rm Pe} \propto  {\rm Pe}^{-2/3}$, which
   will be most
   important for the swimming speed relation (see insets in Fig.\
   \ref{fig:FMtotalPe}). Moreover, these 
   numerical FEM results show that both prefactors in Eq.\
   (\ref{eq:FMtotalMarangoni}) are  of order unity
but hard to quantify because of finite size effects.

This shows that Marangoni flows depress the total driving force
 in the linear response regime by a factor ${\rm Pe}^{-2/3}$
 reflecting the fact that
 it is harder to break
 the symmetry in the presence of the strong Marangoni flow advection.
 Numerical results in Fig.\ \ref{fig:FMtotalPe}
 also show that the total Marangoni force is somewhat larger
 than the direct Marangoni force,
 $\bar{F}_{\rm M,tot} >  \bar{F}_{\rm M}$.
 In this respect,
 our previous results for linear response regime for ${\rm Pe}\ll \bar{U}$
 remain unchanged:
 the Marangoni flow force {\em increases} the direct force.

\begin{figure}
  \begin{center}
    \includegraphics[width=0.99\linewidth]{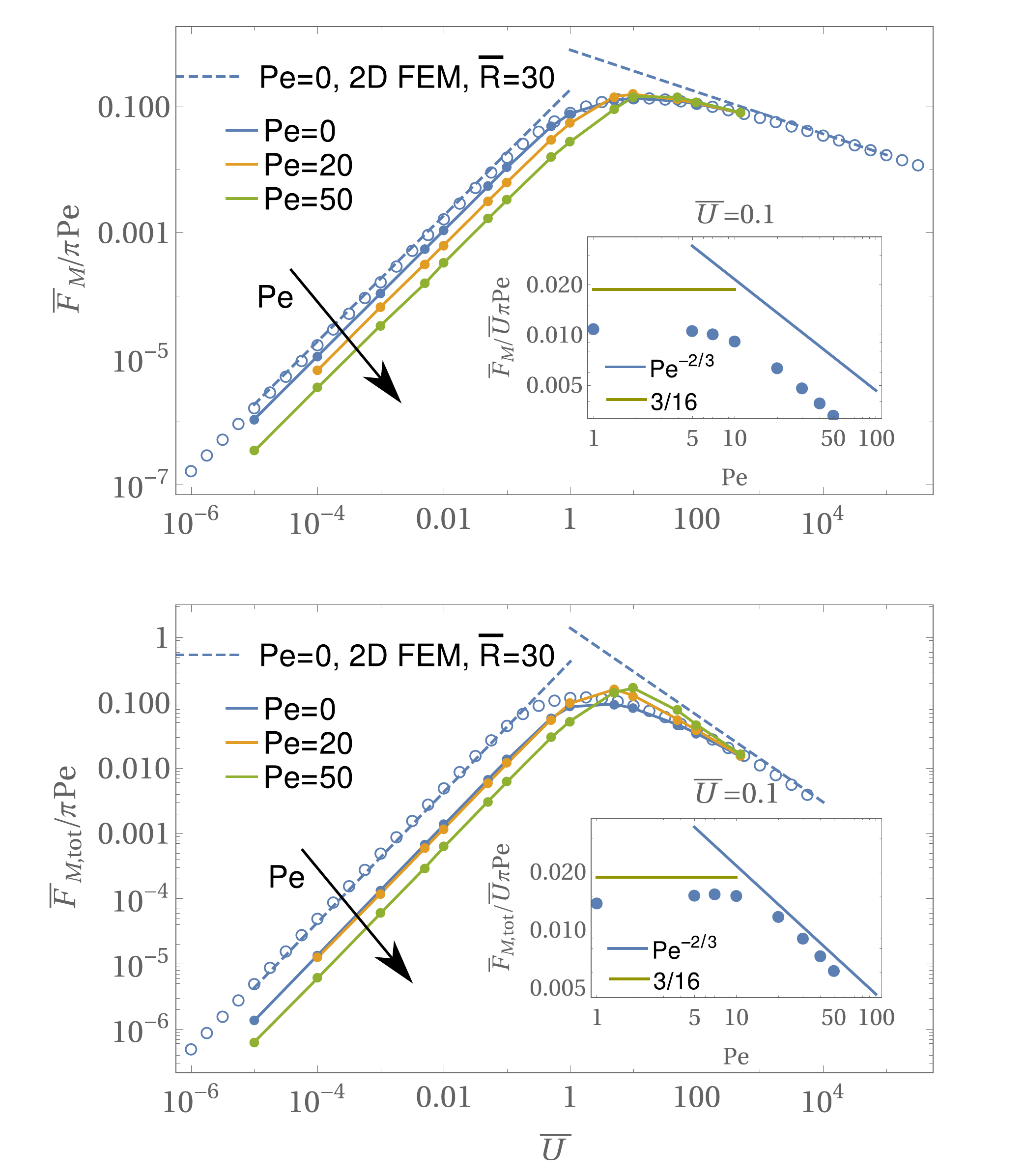}
    \caption{\label{fig:FMtotalPe}
   Iterative three-dimensional  FEM results for
      ${ \bar{F}_{\rm M}}/\pi{\rm Pe}$ (top) and ${\bar{F}_{\rm
        M,tot}}/\pi {\rm Pe}$ (bottom) as a function of $\bar{U}$
    for ${\rm Pe} = 0-50$
    for  a cubic system with  $-7 < \bar{y} <0$,
    $0<\bar{x}<7$, $-7<\bar{z}<7$.
    Insets show the slopes
      ${ \bar{F}_{\rm M}}/\bar{U}\pi{\rm Pe}$ and ${\bar{F}_{\rm
        M,tot}}/\bar{U}\pi {\rm Pe}$  as a function of ${\rm Pe}$
    calculated from the results for  $\bar{U}=0.1$.
     Artificial symmetry breaking from lattice irregularities/defects
    is prevented by averaging all measured quantities
    over two simulations with $U$ and $-U$.
    Blue open circles are results for ${\rm Pe}=0$
    from FEM solutions to the axisymmetric diffusion-advection
    equation in two-dimensional angular
    representation with $\bar{R}=30$.
    The slope in the linear response regime for $\bar{U}\ll 1$
    is reduced according to Eq.\ 
     (\ref{eq:FMtotalMarangoni}). Results for
    $\bar{U}\gg 1$ are essentially not
 affected by strong Marangoni flows ${\rm Pe}\gg {\bar U}$.
  }
\end{center}
\end{figure}

In the advection-dominated regime 
    $\bar{U}\gg 1$, on the other hand, results  are essentially not
    affected by increasing the Peclet number ${\rm Pe}$
      into the regime of strong Marangoni flows ${\rm Pe}\gg {\bar U}$,
    as the numerical results in Fig.\ \ref{fig:FMtotalPe} show:
    all curves for the Marangoni forces converge to the
      previous diffusion-advection results for $\bar{U} \gg 1$.
     The reason for this behavior is that
the flow field $\vec{v}$ will still give
 rise to a concentration boundary layer of thickness
 $\Delta \rho \sim \bar{U}^{-1/3}$
 around the sphere, which determines the  concentration field and,
   thus, the Marangoni forces.
 On the scale of the boundary layer, the Marangoni
 flows $\vec{v}_{\rm M}$ are not yet developed; they develop only 
 further away
 at $1\ll \rho < \rho_{\rm M} \sim  {\rm Pe}$ because of the no-slip
 boundary condition for the Marangoni flow in (iib). 
 Therefore, the results for $\bar{U}\gg 1$ remain essentially 
 unaffected by a strong Marangoni flow for ${\rm Pe}\gg {\bar U}$.

\subsubsection{Evaporation}

In the presence of evaporation,
the boundary condition for the diffusion-advection problem
  changes at the air-water interface $S_{\rm Int}$.
  We then have
 the  convective (Robin)  boundary condition (\ref{eq:bcevap}),
 which is governed by the dimensionless Biot number (\ref{eq:Biot}),
 instead of the Neumann condition (\ref{eq:bcnoevap}),
 which is recovered for vanishing Biot number $\bar{k}=0$.
 In general, evaporation of surfactant
 depletes the interface of surfactant and, thus, decreases
 the Marangoni driving forces (both direct and flow force).

 For volatile camphor, 
 $k \approx \sim 10^{-4}\, {\rm m/s}$
 has been suggested \cite{Soh2008}, which corresponds to 
a high Biot number  $\bar{k} = ak/D \approx 550$ 
 for the camphor disks from \cite{Boniface2019}.
 PEG, on the contrary,
 has a negligible Biot number as it is not volatile. 
As a consequence of the new convective boundary condition, 
 the concentration profile will fall off exponentially
perpendicular to the interface in the outward direction on a
dimensionless extrapolation length scale $\Delta \bar{y}\sim 1/\bar{k}$
given by the inverse of the Biot number.

In the presence of evaporation, we can  develop a qualitative scaling theory
based on the assumption that
the total evaporation flux balances the total emission flux
of surfactant in a stationary state, which gives
\begin{equation}
  -\int_{S_{\rm Int}}   \left. 
 \partial_{\bar{y}} \bar{c}(\vec{\rho})\right|_{\bar{y}=0}
   = 2\pi
\label{eq:jev_balance}
\end{equation}
in dimensionless quantities and  determines the derivatives
  $   \left. 
 \partial_{\bar{y}} \bar{c}(\vec{\rho})\right|_{\bar{y}=0}$
at the scaling level. Via the convective boundary condition
(\ref{eq:bcevap}), this  also determines  the surface concentration
$\left. \bar{c}(\vec{\rho})\right|_{\bar{y}=0}$.
Moreover, the convective boundary condition should reduce
to our previous results for the Neumann condition for small Biot numbers
$\bar{k}$, where the evaporation flux
$\bar{j}_{\rm ev} =\bar{k}\left. \bar{c}(\vec{\rho})\right|_{\bar{y}=0}$
is smaller than the dominating transport flux, which is the diffusive
or Marangoni
flux for  $\bar{U}\ll 1$ and the convective flux for $\bar{U}\gg 1$.

\begin{figure}
  \begin{center}
    \includegraphics[width=0.99\linewidth]{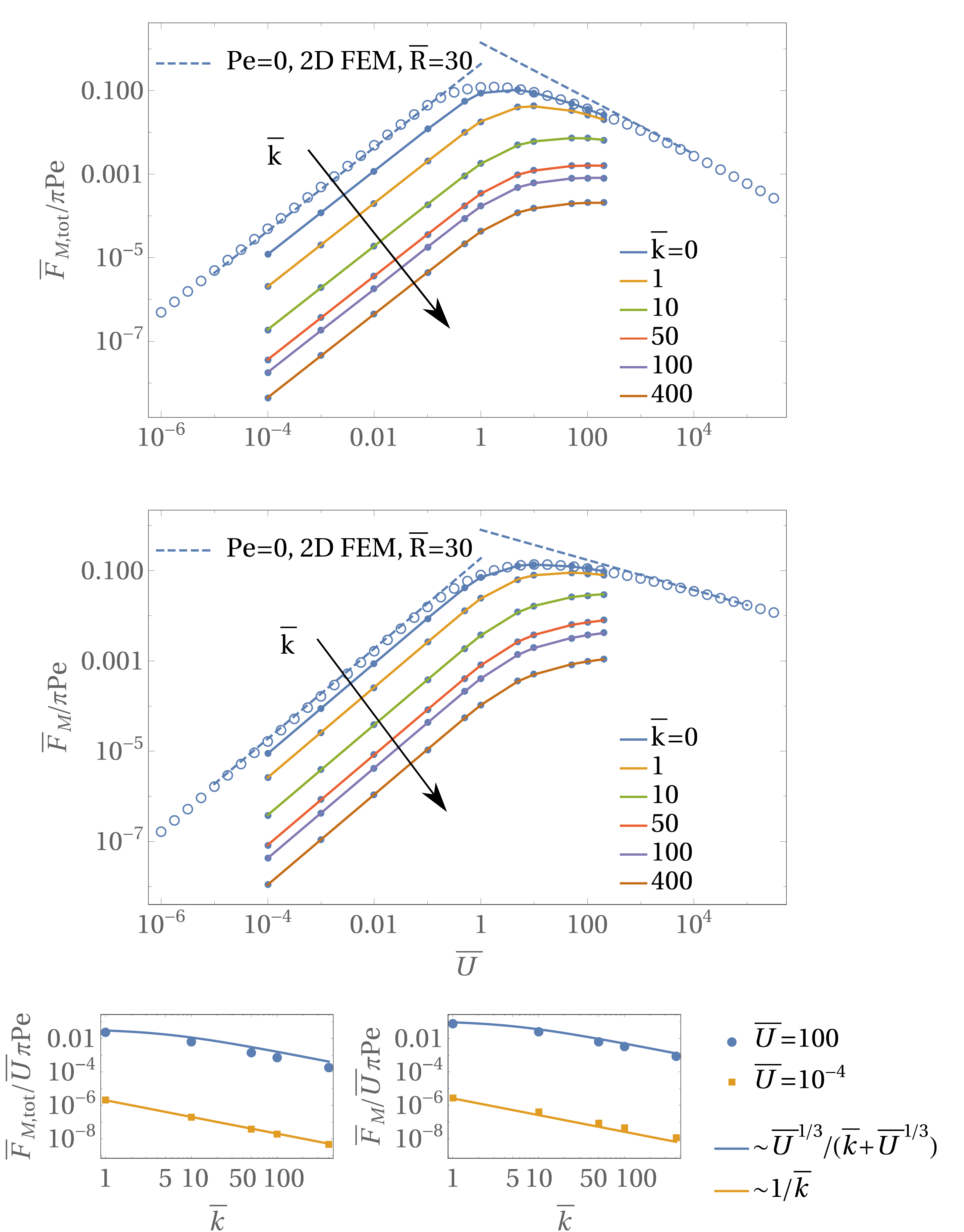}
    \caption{\label{fig:FtotalUk}
 Iterative three-dimensional  FEM results for
      ${ \bar{F}_{\rm M}}/\pi{\rm Pe}$ (top) and ${\bar{F}_{\rm
        M,tot}}/\pi {\rm Pe}$ (middle) as a function of $\bar{U}$
    for ${\rm Pe} = 0$ and Biot numbers $\bar{k}=0-400$
    for a half-cylindrical  system with  $\rho <15$,
    $\bar{x}>0$,      $-4 < \bar{y} <0$,
    Blue dashed lines are results for $\bar{k}=0$
    from FEM solutions to the axisymmetric diffusion-advection
    equation in two-dimensional angular
    representation with $\bar{R}=30$.
    Forces in the diffusive regime  $\bar{U}\ll 1$
    are reduced according to Eq.\
     (\ref{eq:FMtotalkUsmall}). Forces in the advective regime 
     $\bar{U}\gg 1$ are reduced according to  Eq.\ (\ref{eq:FMtotalkU}).
    Bottom row:  FEM results for
    ${ \bar{F}_{\rm M}}/\pi{\rm Pe}$  and ${\bar{F}_{\rm
           M,tot}}/\pi {\rm Pe}$
       as a function of Biot number $\bar{k}$ for $\bar{U}=10^{-4},100$
       in comparison to scaling results in Eqs.\
       (\ref{eq:FMtotalkUsmall}) and  (\ref{eq:FMtotalkU}).
  }
\end{center}
\end{figure}

For the diffusion or Marangoni dominated situation for $\bar{U}\ll 1$,
the concentration and, thus, evaporation is distributed
over the whole interface $S_{\rm Int}$, i.e., there is
  no concentration boundary layer around the half-sphere.
Therefore, flux balance
(\ref{eq:jev_balance}) 
leads to $   \left. 
 \partial_{\bar{y}} \bar{c}(\vec{\rho})\right|_{\bar{y}=0} \sim O(1)$ 
resulting in 
$\left. \bar{c}(\vec{\rho})\right|_{\bar{y}=0} \sim 
1/\bar{k}$
 because of the convective boundary
  condition (\ref{eq:bcevap}).
  Then, also $\bar{F}_{\rm M}\sim 1/\bar{k}$ and
  $\bar{F}_{\rm M,tot}\sim 1/\bar{k}$.
Moreover,   the evaporation flux dominates over the 
diffusive or Marangoni fluxes
(which are  $O(1)$) only for $\bar{k}>\bar{k}_0$ with a 
  crossover value $\bar{k}_0 =O(1)$, which determines the crossover
  to the non-evaporative case. Our numerical results in
  Fig.\ \ref{fig:FtotalUk} suggest $\bar{k}_0 \ll 1$.
Therefore, we expect 
\begin{align}
  \bar{F}_{\rm M}  &\sim
 \left.  \bar{F}_{\rm M}\right|_{\bar{k}=0} \frac{1}{\bar{k}+\bar{k}_0},
&   \bar{F}_{\rm M,tot}
    &\sim  \left. \bar{F}_{\rm M,tot}\right|_{\bar{k}=0}
      \frac{1}{\bar{k}+\bar{k}_0}.
      \label{eq:FMtotalkUsmall}  
 \end{align}
 We checked these predictions numerically in Fig.\ \ref{fig:FtotalUk}
   by using our iterative FEM approach (see Appendix \ref{app:numerical})
   and find good agreement. The  plots in the bottom row  (yellow symbols)
   show that
   the dependence on $\bar{k}$ for $\bar{U}\ll 1$ is described very well by 
   $\bar{F}_{\rm M} =  \left.\bar{F}_{\rm M}\right|_{\bar{k}=1}/\bar{k}$ and
 $\bar{F}_{\rm M,tot} = \left. \bar{F}_{\rm M,tot}\right|_{\bar{k}=1}/\bar{k}$
 suggesting $\bar{k}_0 \ll 1$.

For the advection-dominated situation for $\bar{U}\gg 1$
the concentration and, thus, evaporation is present 
only in the concentration
  boundary layer of radial width  $\Delta \rho \sim \bar{U}^{-1/3}$
  around the half-sphere
and in the advection tail.
 Therefore, the flux balance (\ref{eq:jev_balance}) 
leads to $   \left. 
 \partial_{\bar{y}} \bar{c}(\vec{\rho})\right|_{\bar{y}=0} \sim \bar{U}^{1/3}$ 
and 
$\left. \bar{c}(\vec{\rho})\right|_{\bar{y}=0} \sim 
\bar{U}^{1/3}/\bar{k}$
 because of the convective boundary
  condition (\ref{eq:bcevap}).
  Then, also $\bar{F}_{\rm M}\sim \bar{U}^{1/3}/\bar{k}$
  and $\bar{F}_{\rm M,tot}\sim \bar{U}^{1/3}/\bar{k}$ 
if the evaporation flux dominates over the 
convective fluxes. The convective flux at the interface
and at the boundary layer ($\rho\approx 1+\Delta\rho$)
is in radial direction
$\bar{j}_u =\bar{U}\bar{u}(\rho) \cos\theta  \bar{c}(\rho) \sim
\bar{U}\Delta \rho^2 \bar{c}(\rho)
\sim \bar{U}^{1/3} \bar{c}(\rho)$ and
$\bar{j}_v =\bar{U}\bar{v}(\rho) \sin\theta \bar{c}(\rho) \sim
\bar{U}\Delta \rho \sin\theta \bar{c}(\rho)
\sim \bar{U}^{2/3}\sin\theta \bar{c}(\rho)$ 
in $\theta$-direction. In the advection tail (of angular width
$\Delta \theta \sim
\bar{U}^{-1/3}$), this also leads to $\bar{j}_v \sim \bar{U}^{1/3}
\bar{c}(\rho)$.
Therefore, the evaporation flux starts to dominate over
the convective radial flux and the flux in $\theta$-direction in
the advection tail for $\bar{k}> \bar{U}^{1/3}$; only then we
see the effects of evaporation. 
 Therefore, we expect 
\begin{align}
  \bar{F}_{\rm M}  &\sim
                     \left.  \bar{F}_{\rm M}\right|_{\bar{k}=0}
                     \frac{\bar{U}^{1/3}}{\bar{k}+\bar{U}^{1/3}},
& \bar{F}_{\rm M,tot}
    &\sim  \left. \bar{F}_{\rm M,tot}\right|_{\bar{k}=0}
      \frac{\bar{U}^{1/3}}{\bar{k}+\bar{U}^{1/3}}
      \label{eq:FMtotalkU}
 \end{align}
 for $\bar{U}\gg 1$.
 We checked these predictions numerically in Fig.\ \ref{fig:FtotalUk}
   and find good agreement. The  plots in the bottom row  (blue symbols)
   show that
   the dependence on $\bar{k}$ for $\bar{U}\gg 1$ agrees very well with 
  Eq.\ (\ref{eq:FMtotalkU}).

In summary, we see a reduction of all  Marangoni forces by evaporation
both in the linear response regime $\bar{U}\ll 1$ but also  in the
regime $\bar{U}\gg 1$  of strong advection.  In both regimes,
evaporation reduces the surfactant concentration, which decreases 
the Marangoni forces.

\subsubsection{Force balance and swimming condition}

Now we have a rather
 complete picture of the solution of
problems (i)-(iii), i.e., diffusion-advection coupled to
hydrodynamics 
for a prescribed swimmer velocity $\bar{U}$ at low Reynolds numbers.
The main result is the Marangoni forces as a function of the
prescribed velocity  $\bar{U}$.
The swimming condition (\ref{eq:swimcond0}) or (\ref{eq:swimcond})
gives an additional force balance
relation between Marangoni forces and $\bar{U}$,
which has to be satisfied in the swimming state and determines
the swimming speed $\bar{U}=\bar{U}_{\rm swim}$ as a function
of Peclet number ${\rm Pe}$  and Biot number $\bar{k}$. 
In general, the swimming velocity increases with ${\rm Pe}$ and
decreases with $\bar{k}$.

First, we consider  small ${\rm Pe}$, i.e., small surfactant emission rates
and see whether a swimming state with spontaneously broken symmetry
can exist. 
For $\bar{k}\approx 0$, as appropriate for the PEG-alginate
swimmer, we find from Eq.\ (\ref{eq:c1Mflnew})
that a solution for the swimming condition exists above
a critical Peclet number
${\rm Pe}>  {\rm Pe}_c \sim   8/\ln \bar{R} \to 0$,
which approaches zero for large system sizes. Therefore, the
symmetry is essentially always spontaneously
broken in  a large  swimming vessel.
 Spontaneous symmetry breaking resulting in propulsion
  is  possible by establishing
    an asymmetric surfactant concentration profile that is maintained
    by advection and can produce the necessary Marangoni forces.
Equation  (\ref{eq:c1Mflnew})
is valid only in the decoupled limit ${\rm Pe}\ll \bar{U}$. 
At the swimming bifurcation, we have ${\rm Pe} ={\rm Pe}_c \gg \bar{U}\approx
0$, however,  such that the feedback of  Marangoni flows
onto the diffusion-advection problem has to be taken into account,
 and the decoupling approximation
 cannot  be used. 
 Then, Eq.\ (\ref{eq:FMtotalMarangoni}) describes the Marangoni forces
 in the linear response regime, which further reduces the critical Peclet
 number to ${\rm Pe}_c \sim 1/(\ln \bar{R})^{3}\to 0$.
 In the presence of relevant evaporation $\bar{k}\gg 1$, as
 appropriate for camphor, the total Marangoni force is
 depressed according to Eq.\ (\ref{eq:FMtotalkUsmall}) resulting
 in an increased ${\rm Pe}_c \sim  \bar{k}^3/(\ln \bar{R})^{3} \to 0$, which is,
 however, still approaching zero for large swimming vessel sizes $\bar{R}$.
   Immediate propulsion in all experiments is in accordance with a
   bifurcation with a small ${\rm Pe}_c$. Moreover,
   we observe an intermittent stopping of the swimming motion
   only in the very end (after  $20\, {\rm min}$ or more)
   before 
   the swimming motion stops completely (because the fuel has
   been consumed). This confirms a  small critical value ${\rm Pe}_c$
   below which  ${\rm Pe}$ drops only for very small
   emission current densities $\alpha$.
   Small irregularities can already break the symmetry and give
   rise to an avoided bifurcation and select 
   a fixed swimming axis with respect to the particle orientation,
   which is also observed in the experiments.

 For ${\rm Pe}> {\rm Pe}_c$, a spontaneously symmetry-broken
 swimming state with $\bar{U}_{\rm swim}>0$ exists. 
 Because the Marangoni force  Eq.\ (\ref{eq:c1Mflnew})
 remains approximately linear up to $\bar{U}\sim O(1)$, as can also
 be seen in Fig.\ \ref{fig:FM}, the swimming velocity
 rises steeply for ${\rm Pe}\gtrsim {\rm Pe}_c$ and quickly
 enters the asymptotics for the advection-dominated regime $\bar{U}_{\rm swim}
 \gg 1$. Here, we find  the swimming relations
 \begin{align}
\bar{U}_{\rm swim} &\sim {\rm Pe}^{3/5} 
    &&\mbox{for}~ \bar{k}\ll {\rm Pe}^{1/5},
           \label{eq:PeU}\\
   \bar{U}_{\rm swim} &\sim  \bar{k}^{-3/4} {\rm Pe}^{3/4} 
    &&\mbox{for}~   \bar{k}\gg {\rm Pe}^{1/5}.
           \label{eq:PeUk}
 \end{align}
Also in this regime, we have 
${\rm Pe} \gg \bar{U}_{\rm swim}$ such that
Marangoni flows are strong, but this
has little influence on the swimming speed because of the
concentration boundary layer that forms in the advective regime.
Evaporation is significant  for $\bar{k} \gg {\rm Pe}^{1/5}$ and
reduces the swimming speed 
because it reduces  the driving Marangoni forces. 
 The swimming  relations (\ref{eq:PeU}) and (\ref{eq:PeUk})
  are  shown in Fig.\ \ref{fig:comparison} as dotted yellow and
  dotted blue lines, respectively, together
 with the experimental results for
 our PEG-alginate swimmers and camphor boats
 from Boniface {\it et al.} \cite{Boniface2019}.
 We see clearly, that the experimentally observed swimming speed differs,
 because these swimmers operate at higher Reynolds numbers.

\subsection{High Reynolds numbers}

We have developed a complete picture of the solution of
problems (i)-(iii), i.e., diffusion-advection coupled to
hydrodynamics 
for a prescribed swimmer velocity $\bar{U}$ at low Reynolds numbers,
including evaporation.
Low Reynolds numbers  ${\rm Re} = 2\bar{U}/{\rm Sc} \ll 1$
are realized for $\bar{U} \ll {\rm Sc}/2$, which can still be much larger
than unity for  typical Schmidt numbers for surfactants in aqueous
solutions (see Table \ref{tab:nondimen}).
For the relevant Marangoni propulsion forces,
  the following picture has emerged from our analysis at low Reynolds
  numbers. There is a diffusive regime for $\bar{U}\ll 1$, which
  becomes modified by strong Marangoni flows for
   Peclet numbers ${\rm Pe}\gg \bar{U}$,
 and there is an advective regime
  for $\bar{U}\gg 1$, which is essentially unchanged in the
  presence of strong Marangoni flows for  ${\rm Pe}\gg \bar{U}$
  (see Fig.\ \ref{fig:FMtotalPe}).
  Both regimes are modified in the presence of evaporation if
  the Biot number is $\bar{k} \ge 1$ in the diffusive regime and
  of $\bar{k} \gg \bar{U}^{1/3}$ in the advective regime (see Fig.\
  \ref{fig:FtotalUk}).

  High Reynolds numbers occur for large velocities $\bar{U} \gg  {\rm Sc}/2$
and, therefore, 
always  deep in the advective regime $\bar{U}\gg 1$.
At low Reynolds numbers, the concentration boundary layer of
dimensionless width $\Delta \rho  \sim  \bar{U}^{-1/3}$ (see Eq.\
(\ref{eq:Deltarho})) determines the results for the Marangoni
forces  in this advective limit [see Eqs.\ (\ref{eq:NuU}), (\ref{eq:NuUB}),
(\ref{eq:c1Unew}), (\ref{eq:c1Mflnew}), 
(\ref{eq:FMtotalMarangoni}) and  (\ref{eq:FMtotalkU})].

In order to generalize to higher Reynolds numbers, we realize that
the concentration boundary layer width
is closely related to  the Nusselt number.
By definition (\ref{eq:Nu}),
${\rm Nu} = {-\partial_\rho \bar{c}_0(1)}/{\bar{c}_0(1)}$,
the Nusselt number is an inverse extrapolation length, which
we expect to be the inverse concentration boundary layer width, 
\begin{equation}
  {\rm Nu}  \sim \frac{1}{\Delta \rho}.
  \label{eq:NuDeltarho}
\end{equation}
The result ${\rm Nu} \sim \bar{U}^{1/3}$ from Eqs.\ (\ref{eq:NuU}) and
(\ref{eq:NuUB}) confirms this result both for constant flux and
constant concentration boundary conditions at low Reynolds numbers,
  and we conjecture it to hold also at higher Reynolds numbers.
Phenomenologically, the Nusselt number is well-studied
also for high Reynolds number \cite{Michaelides2003}, both for heat (${\rm
  Nu}_T$ in the following) and for mass transfer (${\rm Nu}$ in the
following, also Sherwood number  ${\rm Sh}$ in the literature),
and we can draw on these results in order to develop
  a theory for the concentration boundary layer and the Marangoni forces.
Up to moderate Reynolds numbers ${\rm Re} \lesssim 200$, the physics
is governed by additional laminar (viscous) boundary layers
that appear around a sphere in fluid flow,
which typically have a width
$ \bar{\delta} \propto {\rm Re}^{-1/2}$
\cite{White2006,Schlichting2016}.
The viscous boundary layer scaling
can be rationalized by generalizing our above
scaling argument for the concentration boundary layer
leading to Eq.\ (\ref{eq:Deltarho}).
The important difference is that the velocity field
  close to the sphere changes from $v(\Delta r) \sim U\Delta r/a$ for Stokes
  flow to $v(\Delta r) \sim U \Delta r/\delta$ for laminar
  boundary layer flow with a no slip boundary condition.
  This leads to 
 \begin{equation}
   \Delta \rho = \Delta r/a  \sim  \bar{\delta}^{1/3} \bar{U}^{-1/3}
   \sim {\bar{U}}^{-1/2} {\rm Sc}^{1/6}.
  \label{eq:Deltarhodelta}
\end{equation} 
This scaling result is in accordance with
phenomenological results for the Nusselt number by 
 Ranz and Marshall \cite{Ranz1952}  (${\rm Re} = 2{\bar{U}}/{\rm Sc}$)
\begin{align}
  {\rm Nu}_T &= 1.0 + 0.3 {\rm Re}^{1/2} {\rm Pr}^{1/3}
               = 1.0 + 0.3 \sqrt{2}\bar{U}^{1/2} {\rm Sc}^{-1/2} {\rm Pr}^{1/3},
   \nonumber\\
  {\rm Nu} &= 1.0 + 0.3 {\rm Re}^{1/2} {\rm Sc}^{1/3}
             = 1.0 + 0.3 \sqrt{2}\bar{U}^{1/2} {\rm Sc}^{-1/6}
 \label{eq:Ranz}
\end{align}
(${\rm Re} = 2{\bar{U}}/{\rm Sc}$ and 
with ${\rm Sc}$ replacing the Prandtl number ${\rm Pr}$ for the mass transfer
Nusselt number).

Because the concentration will drop  in radial direction
  from its value at the surface $S$ of the half-sphere  to zero
  within the concentration
  boundary layer of width $\Delta \rho$, and, thus,
$1 = - \partial_\rho c(\rho=1)  \sim
c(\rho=1)/\Delta \rho$
for constant flux boundary conditions, 
the scaling  of the concentration boundary layer width (\ref{eq:NuDeltarho})
also gives rise to 
\begin{equation}
   \bar{c}(\rho\!=\!1,\theta) \sim {\Delta \rho} \sim {\rm Nu}^{-1},
  \label{eq:cNuDeltarho}
\end{equation}
i.e.,  the symmetry-breaking concentration level at the
sphere is inversely proportional to
the Nusselt number [generalizing Eq.\ (\ref{eq:cDeltarho})]. 
Therefore,   the direct Marangoni force (\ref{FMdim}) should
follow
\begin{equation}
  \frac{\bar{F}_{\rm M}}{\rm Pe} \sim
  \int_0^\pi d\theta \cos\theta  \bar{c}(1,\theta) \sim {\rm Nu}^{-1} \sim
     \bar{U}^{-1/2} {\rm Sc}^{1/6} 
  \label{eq:FMhigh0}
\end{equation}
at higher Reynolds numbers.
The total Marangoni force does no longer follow from a reciprocal
theorem.
In terms of an energy balance, 
the reciprocal theorem can be interpreted as the
absence of mutual dissipation between  swimming flow and Marangoni flow
\cite{Ender2020}.
Therefore,
the power input by surface Marangoni stresses into the swimming flow is,
 transmitted \emph{without loss} as
power input by the Marangoni flow force onto the swimmer. 
For higher Reynolds numbers, the mutual dissipation is no longer zero
but there are additional viscous
terms appearing, which are connected to the vorticity of the flow. 
This suggests that the Marangoni stresses at the interface become less
effective in generating a Marangoni flow force 
because of this  additional dissipation.
Therefore, we simply neglect the Marangoni flow force (or assume that the
 Marangoni flow force is sub-dominant) and
only consider the direct Marangoni
force (\ref{eq:FMhigh0}) at high Reynolds numbers, in the following.

Likewise, the existence of viscous boundary layers around the half-sphere
modifies the drag force.  
On phenomenological grounds,  it has been suggested  that
 $F_{\rm D}  = D_c \frac{\pi}{2}\mu a U$ with
 $D_c\simeq 6{\rm Nu}_T$ \cite{Duan2015},
where ${\rm Nu}_T$ is the  Nusselt number for heat transport,
resulting in
\begin{equation}
  \bar{F}_{\rm D} = -3\pi \bar{U} {\rm Nu}_T.
  \label{eq:FShigh}
\end{equation}
Using the Ranz and Marshall correlation (\ref{eq:Ranz}), we find from
the force balance  $\bar{F}_{\rm D} +\bar{F}_{\rm M}=0$
\begin{align}
  {\rm Pe}&\approx 3 {\rm Nu}{\rm Nu}_T  \bar{U}_{\rm swim},
            \nonumber\\
  \bar{U}_{\rm swim} &\sim     {\rm Sc}^{1/3} {\rm Pr}^{-1/6}
                       {\rm Pe}^{1/2}
           \label{eq:PeUhigh}           
\end{align}
in the absence of evaporation.
In the presence of evaporation, we use
$\bar{F}_{\rm M} \sim \left.\bar{F}_{\rm M}\right|_{\bar{k}=0}
{{\rm Nu}}/{(\bar{k}+{\rm Nu})}$
[cf.\ Eq.\ (\ref{eq:FMtotalkU})] to find
\begin{align}
  {\rm Pe}&\approx 3 (\bar{k}+{\rm Nu}){\rm Nu}_T  \bar{U}_{\rm swim},
            \nonumber\\
    \bar{U}_{\rm swim}     &\sim    \bar{k}^{-2/3} 
  {\rm Sc}^{1/3} {\rm Pr}^{-1/2} {\rm Pe}^{2/3}.
  \label{eq:PeUhighk}
\end{align}
Both results (\ref{eq:PeUhigh}) and (\ref{eq:PeUhighk})
are also shown in Fig.\ \ref{fig:comparison}
together with the experimental data on PEG-alginate  and camphor
  Marangoni boats.

\section{Comparison with experiment}

\begin{figure}
  \begin{center}
  \includegraphics[width=0.99\linewidth]{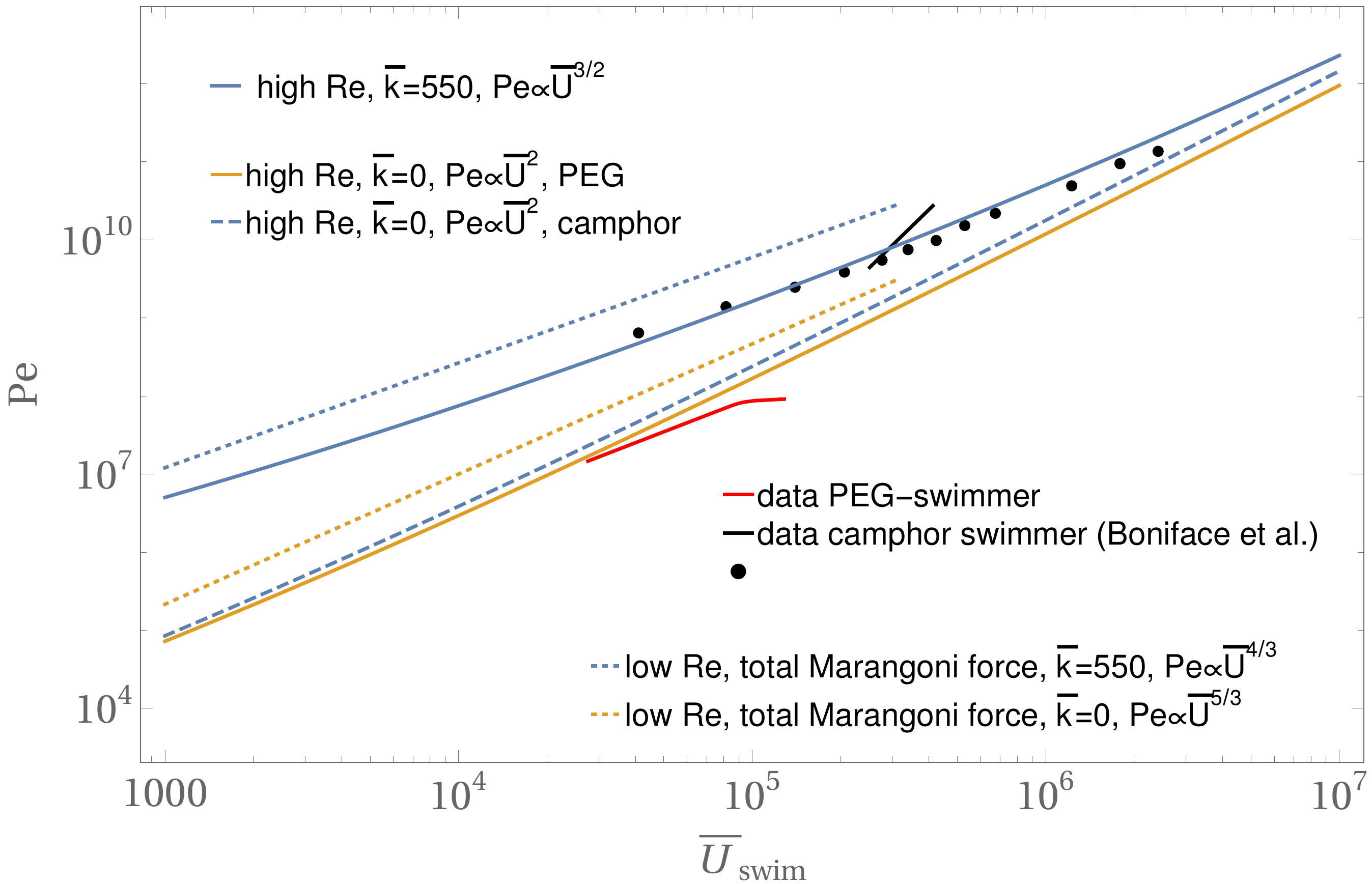}
  \caption{Different theory results for
     ${\rm Pe}$-$\bar{U}_{\rm swim}$ swimming relations in a
    double-logarithmic plot in comparison to  experimental results 
    on PEG-alginate swimmers (red lines, obtained from the data in Fig.\
    \ref{fig:mdot}) and camphor boats
    from Ref.\ \cite{Boniface2019} 
      (black lines and dots; black line
      is time-dependent data, black
     dots are data at fixed times but for different
    radii $a$).
    PEG-alginate swimmers are described well
    by the high Reynolds number theory
    without evaporation (Biot number $\bar{k}=0$);
    camphor boats are best described
    by a high Reynolds number theory with
    Biot number $\bar{k}\approx 550$.
    The corresponding low Reynolds number theories give significantly
      lower swimming speeds. 
  }
   \label{fig:comparison}
   \end{center}
\end{figure}

Force balance for low and high Reynolds numbers
results in a characteristic
${\rm Pe}$-$\bar{U}_{\rm swim}$-relation for the swimmer
with characteristic power laws. 
For low Reynolds numbers, these are relations (\ref{eq:PeU}),
$\bar{U}_{\rm swim} \propto  {\rm Pe}^{3/5}$,
without evaporation 
and (\ref{eq:PeUk}),
$\bar{U}_{\rm swim} \propto  {\rm Pe}^{3/4}$,
in the presence of strong evaporation. 
For higher Reynolds numbers, we find 
relations (\ref{eq:PeUhigh}),  $\bar{U}_{\rm swim} \propto  {\rm Pe}^{1/2}$
without evaporation 
and  (\ref{eq:PeUhighk}), $\bar{U}_{\rm swim} \propto {\rm Pe}^{2/3}$
for strong evaporation.
Also, the experiment on PEG-alginate swimmers and  the camphor boats
from Boniface {\it et al.} \cite{Boniface2019}  take place
at higher Reynolds numbers; parameter value estimates
for these experiments are summarized in   Table \ref{tab:experiment},
the resulting dimensionless parameters in Table \ref{tab:nondimen}.
  Our experimental results for the mass release mass
  $\dot{m}(t)$ as a function of time
(see Fig.\ \ref{fig:mdot}(middle)) and the corresponding  swimming velocity 
$U_{\rm swim}(t)$ (see Fig.\ \ref{fig:mdot}(right)) of the
PEG-alginate swimmers
give the red line in Fig.\  \ref{fig:comparison} in the
${\rm Pe}$-$\bar{U}_{\rm swim}$ parameter plane.
 We also show  experimental results for
 camphor boats from Boniface {\it et al.} \cite{Boniface2019} (black data
   points from experiments varying the radius and black line from
   time-dependent swimming data). 
 Figure    \ref{fig:comparison} compares these experimental
 results with  our
theoretical results for the appropriate parameter
values for the PEG-alginate swimmers and the camphor boats
from  Tables \ref{tab:nondimen} and \ref{tab:experiment}.

\begin{table*}
  \begin{center}
  \caption{\label{tab:experiment} Estimates of experimental parameters.}
     \begin{tabular}{ l | l | l}
     \hline\noalign{\smallskip}
     Parameter &          PEG-alginate swimmer
     & Camphor   boat\cite{Boniface2019}
     \\
   \noalign{\smallskip}\hline\noalign{\smallskip}
     Radius   $a$ &  $1500 \,{\rm \mu m}$  &     $1000-15000\, {\rm \mu m}$ \\
       Diffusion constant  $D$  & $350  \,{\rm \mu m^2/s}$
            & $720 \,{\rm \mu m^2/s}$\\    
 Surface tension reduction  $\kappa = - {\Delta \gamma}/{\Delta c}$
            &
              $2.7\times 10^{-4} \,{\rm m^3/s^2}$   &
                          $2\times 10^{-2} \,{\rm m^3/s^2}$ \\
   Mass loss per time   $\dot{m}= c\pi a^2\alpha$ &
             $c_{\rm half-sph}=2$
  &
     $c_{\rm disk}=1$
       \\
        &
          $0.01-0.25 \times 10^{-6} \,{\rm g/s}$
  &
      $(0.76 \times  10^{-6} \, {\rm g/s} )\left({a}/{4{\rm mm}}\right)^2$
     \\
\noalign{\smallskip}\hline
     \end{tabular}
     \end{center}
\end{table*}

The swimming relations in Fig.\  \ref{fig:comparison} are
the main result of the paper.
For the PEG-alginate swimmer (red line) we see good agreement between
the high Reynolds number theory in the absence of evaporation,
i.e., with  Biot number of $\bar{k} =0$ (yellow line). The corresponding
low Reynolds number theory (dotted yellow line)  gives significantly
lower swimming velocities at the same Peclet number. This outcome
is what we expected based on the above estimate of
moderate Reynolds numbers 
${\rm Re} \sim 60$ for the PEG-alginate swimmers and based on the
non-volatility of PEG.
  We also see that the  slower second phase
  of the swimming motion of the PEG-alginate swimmers
  is described slightly better by our theory (left part of the
  red line in Fig.\  \ref{fig:comparison}), which is in accordance
  with our initial observation that the time constants for mass release and
  swimming velocity agree only in the second phase.
  In the first phase, the Peclet number
$\text{Pe}$ necessary to achieve the measured swimming velocity
is   slightly \emph{lower} than predicted by our
theory, i.e., a more efficient propulsion.
  This is a hint that some of our theoretical assumptions could be
  violated during the  first phase, for example, regarding the
  adsorption equilibrium, which might not yet be established
  starting from an initially ``empty'' air-water interface,
  which could give rise to steeper concentration gradients
  and more efficient propulsion.

For volatile camphor, a Biot number of $\bar{k} \approx 550$
has been suggested in Ref.\  \cite{Soh2008}, which we
use in Fig.\ \ref{fig:comparison} to compare with the
experimental data of Boniface {\it et al.} on camphor disks
\cite{Boniface2019}  (black data points and black line).
Again, we obtain good agreement
with the high Reynolds number theory (blue line).
The disk geometry differs from the half-spherical geometry
we discussed in detail,  but we expect that
the swimming relation will only differ  by numerical factors of order unity. 
The corresponding low Reynolds number theory (dotted blue line)
significantly
underestimates swimming velocities, and a theory without evaporation
(dashed blue line) overestimates swimming velocities.

\section{Discussion and conclusion}

We presented an experimental realization of alginate capsule
self-propulsion at the air-water interface
by loading the alginate capsule with surfactant molecules during
synthesis. Self-propulsion of these capsule swimmers is
based on a Marangoni boat mechanism. 
Alginate is bio-compatible and widely used for capsule production,
which are interesting aspects for further applications. 
The  versatile and simple synthesis
strategy allowed us to identify  various substances
that can propel alginate capsules, see Table \ref{tab:fuels}.
PEG surfactants exhibit the best propulsion properties: for PEG-300,
we find a fast and sustained motion with swimming speeds 
$U_{\rm swim} \sim 2-3 \,{\rm cm/s}$ over $20\, {\rm min}$ and more.
The swimming speed corresponds to several swimmer diameters
per second and is comparable or superior to other self-phoretic
or microswimmers
\cite{ebbens2010} or active liquid droplets \cite{Herminghaus2014}.
In general, we find prolonged propulsion only if 
spreading molecules are water-soluble as the PEG molecules are; then the 
air-water interface can regenerate by the fuel being dissolved
in water. Evaporation from the air-water interface is another mechanism
  to achieve  regeneration, which is utilized in camphor
  boats \cite{Soh2008,Suematsu2014,Akella2018,Boniface2019}.
  We conclude that a mechanism that regenerates
  the air-water interface, such as
  water-solubility or evaporation of surfactants,
  is crucial for prolonged propulsion.

  We could produce alginate swimmers
   down to radii of several hundreds of micrometers,
which is slightly above the realm of low Reynolds numbers.
The future work could address further miniaturization of capsules.

Starting from  low Reynolds numbers, 
we developed a theory for  Marangoni boat propulsion of 
a completely symmetric,  half-spherical, surfactant emitting
swimmer. The theoretical description comprises the coupled
problems of surface tension reduction by surfactant adsorption
at the air-water interface including the possibility of surfactant
evaporation, fluid flow (both Marangoni flow and
flow induced by swimmer motion), diffusion and advection of the
surfactant. In particular, advection is systematically
  included in our approach and turns out to be essential
  for all swimmer velocities $U \gg D/a$ ($\bar{U} \gg 1$).
  These three problems are first solved for prescribed
swimmer velocity $U$; the actual swimming velocity
$U_{\rm swim}$ is determined by force balance between the
drag force, the direct Marangoni force from the surface tension
contribution at the  air-water-swimmer   contact line, and
the Marangoni flow force.
We find that Marangoni
   flows can either act to increase the direct Marangoni force (at 
 low velocities) or to  increase the drag (at
 higher velocities).
 For low Reynolds numbers,
 all theoretical results are supported by numerical FEM simulations.
Non-dimensionalization shows that the swimmer is controlled by
two dimensionless control parameters, the Peclet number (\ref{eq:Pe}),
which is the dimensionless emission rate of surfactant, and
the Biot number (\ref{eq:Biot}), which is the dimensionless
evaporation rate.  Evaporation is practically absent for PEG, but strong
for other frequently studied Marangoni boat swimmers, such as camphor
boats \cite{Soh2008}.

We showed that a spontaneous symmetry breaking, i.e., a spontaneous
transition into a swimming state is possible also for a completely
symmetric swimmer above a critical Peclet number.
Spontaneous symmetry breaking resulting in propulsion
  is  possible    by establishing
    an asymmetric surfactant concentration profile that is maintained
    by advection.
    We find that  the critical Peclet number for this transition
    approaches zero  logarithmically  for large system sizes,
     ${\rm Pe}_c \propto  1/(\ln \bar{R})^{3}$.
The possibility of such a spontaneous symmetry breaking has been
pointed out for  autophoretic swimmers
\cite{michelin2013,Michelin2014}
and liquid Marangoni swimmers \cite{Izri2014} before.
Also in these systems, advection by the surrounding fluid can maintain
the necessary concentration gradients in fields and/or concentrations.

In Eqs.\ (\ref{eq:PeU}) and (\ref{eq:PeUk}), we obtain the power laws
governing the swimming velocity as a function of Peclet and Biot number,
which are $\bar{U}_{\rm swim} \propto  {\rm Pe}^{3/5}$,
without evaporation (PEG)  and
$\bar{U}_{\rm swim} \propto \bar{k}^{-3/4} {\rm Pe}^{3/4}$,
in the presence of strong evaporation (camphor).
This demonstrates that additional evaporation  reduces
swimming speed. 

Experimentally realizable  PEG-alginate or camphor swimmers
are operating at moderate Reynolds numbers around 60 or more. 
Accordingly, we generalized the theoretical approach to
higher Reynolds numbers by using the concept of the Nusselt number,
for which many results at higher Reynolds numbers
are known phenomenologically.
This might also account for some effects related to the formation
of vortices  around the swimmer during propulsion at
higher Reynolds numbers (see PIV-results in Fig.\ \ref{fig:PIV} and
Ref.\ \cite{Sur2019}). 
Finally, we obtained the  swimming relations
(\ref{eq:PeUhigh}) and (\ref{eq:PeUhighk}), which
give $\bar{U}_{\rm swim} \propto  {\rm Pe}^{3/4}$,
without evaporation (PEG)  and
$\bar{U}_{\rm swim} \propto \bar{k}^{-2/3} {\rm Pe}^{1/2}$,
in the presence of strong evaporation (camphor).
We find a good quantitative fit (without any free fitting parameters)
with our own experimental results on PEG-alginate swimmers and
the results of Ref.\ \cite{Boniface2019} on camphor swimmers
in Fig.\ \ref{fig:comparison}.
This is the main result of this paper.
The future work should extend the numerical approach to higher
  Reynolds numbers in order to verify our scaling
  results for the swimming relation using, for example, the methods
  introduced in Ref.\ \cite{JafariKang2020}.
There are several aspects of the self-propulsion of
PEG-alginate capsules, where we
presented  first experimental results but which deserve
a much more detailed investigation in future work: curved trajectories,
interactions with container walls, and swimmer-swimmer interaction.

Curved trajectories as observed in Figs.\ \ref{fig:trajectory} and
\ref{fig:trajectory2} with a swimming direction of the swimmer that
is, at least, weakly linked to its orientation, while the orientation
of the swimmer is slowly turning can only be explained by
small asymmetries of capsules induced by irregularities
in the pore structure. This view is supported by
the individual character of the turning characteristics of
different swimmers (see Fig.\ \ref{fig:trajectory2}).
Future work should explore the relation between capsule irregularities
and turning statistics in more detail.
Experimentally, purely rotary systems could be constructed \cite{Koyano2017}.

PEG-alginate swimmers are repelled by walls. In normal collisions we observe
 direction reversal without reorientation of the swimmer. 
In the framework of the Marangoni boat mechanism, this can be explained by 
 an accumulation of surfactant emitted by the
 swimmer in front of the wall because of the zero flux boundary
 condition at the wall. Surfactant accumulation creates a gradient in 
 surfactant concentration toward the wall, and the swimmer
 reverses direction if advection and accumulation balance without
 changing its  orientation.
 This behavior is similar to what has been observed for asymmetric
 \cite{Hayashima2001} and symmetric \cite{Nagayama2004}
 camphor boats \cite{Nakata2015}.
During the collision the orientation of the swimmer particle
  does not change,
  while the  swimming direction reverses; therefore, the swimming direction
  also reverses relatively to the particle orientation.
  This is consistent with
a weak symmetry breaking by 
small  irregularities in the pore
distribution, which give rise to  many
possible metastable propulsion directions.
A perturbation as during surfactant accumulation and 
direction reversal at the wall can easily cause a change
between these propulsion directions. 
  There are more oblique collisions  (see Fig.\ \ref{fig:trajectory2}),
  which take longer  and can feature a reorientation of the swimmer.
  The underlying mechanisms could be similar to the reorientation
  mechanisms of
self-diffusiophoretic swimmers  \cite{Uspal2015,Bayati2019}
but this issue also requires future work.

  Finally, we have experimental evidence  that PEG-alginate
  swimmers interact with each other
  via their surfactant concentration fields.
  Similar observations have been made already in Refs.\
 \cite{Kohira2001,Heisler2012,Nakata2015}, mostly in channel geometries. 
 We monitored different collisions between swimmers, where
 we could find both attraction and repulsion. In the framework of
 the Marangoni boat mechanisms, where swimmers prefer to move opposite
 to surfactant concentration gradients, we expect that swimmers
 are repelled by their surfactant tails, which represent traces
 of high concentration. This predicts
 a kind of ``chemo-repellent'' behavior with respect to the tails.
 For the interaction between swimmers, the concentration dependence
 of the surface activity $\kappa$ [the $c_0$-dependence
 in Eq.\ (\ref{eq:kappa})]
 can also play an important role \cite{Heisler2012}.
 The topic of swimmer interactions appears to be very rich and important
 for applications regarding the swarming of PEG-alginate
 swimmers; it deserves a much more detailed
 investigation in future work.

\begin{acknowledgement}
We acknowledge financial support by the Deutsche Forschungsgemeinschaft 
via SPP 1726 ``Microswimmers'' (KI 662/7-1,  KI 662/7-2, and Re 681/25-1).
We thank Monika Meuris and the ZEMM (Zentrum f{\"u}r Elektronenmikroskopie
und Materialforschung), TU Dortmund for providing SEM images,
and Peter Ehrhard (Department of Biochemical and Chemical Engineering,
Fluid Mechanics) for access to PIV measurements.  
\end{acknowledgement}

\section{Authors contributions}
H.R. and A.-K.F.  conceived  the experiments.
A.-K.F. performed the experiments. J.K. and H.E.
developed the theoretical model and 
performed the analytic calculations and  numerical
simulations. J.K. wrote the manuscript with support from H.E. and H.R.
All authors discussed the results and commented on the manuscript.
%

\appendix
\numberwithin{equation}{section}

\section{Numerical methods}
\label{app:numerical}

For  ${\rm Pe} \ll \bar{U}$, the Marangoni flow problem decouples
and the advection problem becomes axisymmetric. Then, $c=c(\rho,\theta)$
only depends on the radial coordinate and one angular coordinate and
we solve the diffusion-advection problem on a
two-dimensional rectangular domain in the $\rho$-$\cos\theta$ plane.
Typically we use $\rho< 30$ and FEM-routines from Wolfram MATHEMATICA;
we use irregular
triangular meshes with a  mean  area of mesh elements of $0.00015$
in  the  $\rho$-$\cos\theta$ plane (the maximal area is
$0.005$). We checked that  discretization effects are negligible.

For  ${\rm Pe} \gg \bar{U}$, we have to solve the coupled problems
of Marangoni flow field (iib) and  diffusion-advection 
equation (iii) with the appropriate boundary conditions
from the adsorption problem (i).
Numerically, we only address low Reynolds numbers for the fluid
flow such that the Stokes equation applies.
For the coupled problem we
use an iterative scheme of three-dimensional FEM solutions
both for Marangoni flow and for diffusion-advection,
again employing FEM-routines from Wolfram
MATHEMATICA. For the Stokes flow, we use the analytical result
[Eqs.\ (\ref{u}) and (\ref{v})]. 
In the iterative scheme we  solve for the Marangoni flow field
(iib) starting from an  initial guess for the
concentration profile (typically $\bar{c}^{(0)}(\rho) =1/\rho$).
Then, we use this Marangoni flow field
in the advecting fluid flow field in  the FEM
solution $\bar{\vec{v}}_{\rm M}^{(1)}(\vec{\rho})$ of the diffusion-advection 
equation (iii), which gives in turn 
an improved approximation $\bar{c}^{(1)}(\vec{\rho})$
for the concentration profile.
With this improved approximation we go back into solving for the 
Marangoni flow field (iib) to obtain the next iteration
$\bar{\vec{v}}_{\rm M}^{(2)}(\vec{\rho})$, and so on.
This results in improving approximations $\bar{c}^{(n)}(\vec{\rho})$
and $\bar{\vec{v}}_{\rm M}^{(n)}(\vec{\rho})$, and the iteration
 typically  converges within $n=5-10$ iterations
to the final Marangoni flow field and surfactant concentration field.
The iterative approach has the advantage that the Marangoni boundary
condition in the fluid flow problem (iib) is a fixed one
at each iterative step
and only adjusts over the iteration;
the coupling of the two problems is correctly established over the
iteration. Similar iterative numerical schemes for coupled problems have been
applied successfully in Refs.\ \cite{Boltz2015,Wischnewski2018}.
The iterative scheme works well up to  ${\rm Pe}\sim 50$ for the
system sizes we use. As outlined in the section
on strong Marangoni flow in the limit ${\rm Pe} \gg \bar{U}$,
Marangoni flows are fully established  on the scale
$\rho_{\rm M} \sim  {\rm Pe}$, which is growing with the Peclet number.
Therefore, numerical problems typically arise for higher Peclet numbers
when the   Marangoni roll becomes strongly distorted by the system
boundaries, which gives rise to numerical
instabilities and a failure of the iteration procedure.
This is why the data presented in Fig.\ \ref{fig:FMtotalPe}
is limited to the range ${\rm Pe} = 0-50$.

The FEM solution  of the stationary equations (iib) and (iii)
is obtained on a cylindrical or cubical irregular tetrahedral mesh.
We use  cubical volumes (for example, with edge length 14
in $\bar{x}\bar{z}$-plane  and height 7 in $\bar{y}$-direction
in Fig.\ \ref{fig:FMtotalPe}) for the FEM calculations.
Cylindrical volumes can also be easily implemented.
   The maximal volume of mesh elements is $0.2$,  the mean volume is
   $0.01$.
    Mesh volumes are smaller ($<0.005$)
     in the region $-1<\bar{y}<0$ below the
     interface to capture Marangoni advection.
     Again, we checked that discretization effects are small.
Because of the mirror symmetry $\bar{x}\to -\bar{x}$, we only need to solve on
half-cubes $\bar{x}>0$ and apply
Neumann boundary conditions $\left. \partial_{\bar{x}}
  \bar{c}\right|_{\bar{x}=0}=0$ and $\left. \partial_{\bar{x}}
  \bar{\vec{v}}_{\rm M}\right|_{\bar{x}=0}=0$ to  enforce the mirror symmetry. 
The boundary conditions at the outer boundaries are Dirichlet
conditions 
for the concentration $\bar{c}=0$ and the Marangoni flow
$\bar{\vec{v}}_{\rm M}=0$. For sufficiently large  cubes or cylinders
these boundary conditions should not matter but we still have finite
size effects, in particular at larger Peclet numbers ${\rm Pe}>50$
as explained above. In particular,
our systems are always several particle radii long in every
direction such that strong confinement effects as observed
in Ref.\ \cite{JafariKang2020} for system sizes below two particle radii
should be absent.

We are interested in the resulting symmetry-breaking Marangoni forces
caused by a symmetry-breaking swimming motion as a function
of the  velocity $\bar{U}$.
At small $\bar{U}$ there is the problem that 
artificial symmetry breaking from lattice irregularities/defects
is often larger than symmetry breaking by swimming.
Therefore, we  average all measured quantities
over two simulations with $\bar{U}$ and $-\bar{U}$ to
cancel artificial  symmetry-breaking  effects. 
This step is crucial to obtain accurate results.

\bibliographystyle{epj}
\bibliography{marangoni2}

\end{document}